\documentclass[11pt]{article}
\usepackage{tikz}
\usepackage{geometry}
\usepackage[english]{babel}
\usepackage[utf8]{inputenc}
\usepackage[T1]{fontenc}
\usepackage{indentfirst}
\usepackage{amsmath}
\usepackage{amssymb}
\usepackage{amsthm}
\usepackage{bm}
\usepackage{physics}
\usepackage{proof}
\usepackage{eufrak}
\usepackage{graphicx}
\usepackage{mathtools}
\usepackage{psfrag}
\newcommand{\veff}{v_{\text{eff}}}
\newcommand{\dr}{\text{dr}}
\newcommand{\dperdk}{\frac{d}{d\kappa}}

\allowhyphens

\newcommand{\FQa}{\mathcal{Q}_\alpha}
\newcommand{\FJa}{\mathcal{J}_\alpha}

\newcommand{\edr}{e^{\text{DR}}}
\newcommand{\tdl}{\lim_{\text{TDL}}}
\newcommand{\complex}{\mathbb{C}}
\newcommand{\egesz}{\mathbb{Z}}

\newcommand{\valos}{\mathbb{R}}
\newcommand{\eps}{\varepsilon}

\newcommand{\ordo}{\mathcal{O}}
\newcommand{\lan}{{\boldsymbol \lambda}_N}

\newcommand{\mun}{{\boldsymbol \mu}_N}
\newcommand{\mRR}{\Theta}

\newcommand{\vevv}[1]{\left\langle #1 \right\rangle}


\usepackage{ifpdf}

\ifpdf
\usepackage{epstopdf}
\usepackage[pdftex,colorlinks,urlcolor=blue,citecolor=blue,linkcolor=blue]{hyperref}
\else
\usepackage[hypertex,colorlinks,urlcolor=blue,citecolor=blue,linkcolor=blue]{hyperref}
\fi
\begin{document}
\numberwithin{equation}{section}

\title{Current operators in integrable models: A review}
\author{M\'arton Borsi$^{1,2}$, Bal\'azs Pozsgay$^{1,2}$, Levente
  Pristy\'ak$^{2,3}$}
\date{${}^1$ {\it
  Department of Theoretical Physics, \\
  E\"otv\"os Lor\'and University Budapest
}\\
${}^2$ {\it
  MTA-ELTE ``Momentum'' Integrable Quantum Dynamics Research Group,\\
  E\"otv\"os Lor\'and University Budapest
}\\
${}^3$ {\it Department of Theoretical Physics, \\
  Budapest University of Technology and Economics\\
}
}

\maketitle
\abstract{We consider the current operators of one dimensional
  integrable models. These operators describe the flow of the
  conserved charges of the models, and they play a central role in
  Generalized Hydrodynamics. We present the key statements about the
  mean currents in finite volume and in the thermodynamic limit, and
  we review the various proofs of the exact formulas. 
  We also present a few new results in this review. New contributions include a computation of the
  currents of the Heisenberg spin chains using the string hypothesis,
  and simplified formulas in the thermodynamic limit. We also discuss
  implications of our results for the asymptotic behaviour of
  dynamical correlation functions.}

\section{Introduction}

One dimensional integrable models are special many body systems whose exact solution is possible with analytic
methods. A common property of integrable models is the existence of a large number of 
extra conservation laws 
\cite{caux-integrability}. In classical models 
the conserved quantities are functions on the phase space, and they have vanishing Poisson bracket.
In quantum mechanics the conserved charges are described by Hermitian operators that commute with
each other. In this work we treat quantum mechanical models and we put a special emphasis
on lattice models, in particular on spin chains.

The existence of the extra conservation laws has dramatic effects on the dynamics of these models.
First of all, it can be argued that the scattering events are always
elastic, there is no momentum transfer, and thus no dissipation into
the low energy modes in these systems. Furthermore, the elastic
$S$-matrix factorizes: it is 
always a product of two-body $S$-matrices, and it does not depend on the particular
temporal ordering of the scattering events. These particular physical properties underlie the
solvability of these systems \cite{sutherland-book,Mussardo-review}.

The existence of the commuting set of the charges has been known since a long time,
however, many important consequences were only discovered and understood in the last 5-10 years. One
such consequence is the absence of thermalization in integrable
models: it is now understood that
they
equilibrate to steady states that are described by the so-called Generalized Gibbs Ensemble
(GGE) \cite{essler-fagotti-quench-review}. Furthermore, the large scale transport properties of the
models can be described by focusing 
on the dynamics of the conserved charges only: this led to the Generalized Hydrodynamics (GHD)
\cite{doyon-ghd,jacopo-ghd}, which is the topic of this special issue.

The construction of the conserved charges and the computation of their eigenvalues is well
understood for an important class of integrable models: the
Yang-Baxter integrable spin chains. These cases
can be treated by the Quantum Inverse Scattering Method
(QISM) developed by L. Faddeev and the Leningrad group \cite{faddeev-history,Korepin-Book}.
The charges are
obtained from a one-parameter family of commuting transfer matrices:  they are given as expansion
coefficients around special points. Typically the Hamiltonian is the first or second expansion
coefficient and it involves nearest neighbor interactions. It is important that there are other types of integrable spin
chains where the construction of the charges is much more involved: these are long
range interacting models such as the Haldane-Shastry model \cite{haldane-shastry-conserved} or the
Inozemtsev spin chain \cite{inozemtsev-chain}.

In contrast to the charges, much less was known about the current
operators, which describe the flow of conserved quantities under time
evolution generated by the Hamiltonian.
In field theories with Lorentz invariance the charge densities and the currents are
intimately related: they are components of a Lorentz
2-vector. However, in spin chains and in non-relativistic 
continuum models the currents are fundamentally different from the
charges.

The motivation to study the
current operators came mainly from GHD: the
central equation of GHD for the ballistic transport is a continuity
equation that involves the mean currents in a local equilibrium. It
was stated in  \cite{doyon-ghd,jacopo-ghd} that the mean currents can
be computed by an essentially semi-classical formula, which
counts the charge eigenvalues carried by the fundamental
excitations, where a semi-classical propagation speed is computed
from the exact quantum mechanical solution of the models. These 
statements for the current mean values triggered research activity in
the last 4 years, which contributed significantly to the understanding
of current operators in integrable models. It is the goal of this work
to review the recent results.

The paper is structured as follows. In Sec. \ref{sec:tdlintro} we
review the key statements about the current mean values in the
thermodynamic limit; these statements  were first introduced in
the seminal works \cite{doyon-ghd,jacopo-ghd}. It will be the topic of the later
Sections to discuss the various proofs and implications of these
statements.  In \ref{sec:main} we introduce formulas for the finite
volume mean values of the currents, which are proven in later
Sections. In \ref{sec:tdl} we show that the original statements of
\cite{doyon-ghd,jacopo-ghd} indeed follow from the finite volume
formulas; this step has not yet been discussed in the literature.
Section \ref{sec:xxx} is devoted to the current mean values of the XXX
Heisenberg spin chain. Sections \ref{sec:fftcsa2}, \ref{sec:algebraic}
and \ref{sec:longrange} include three different proofs of the finite
volume current formulas, with \ref{sec:algebraic} also discussing a
new algebraic construction, which embeds the current operators into the QISM
framework. In Section \ref{sec:corr} we discuss implications to the
theory of factorized correlation functions. We conclude in Sec. \ref{sec:conclusions} and present a list of open
questions. 

\section{Current mean values in the thermodynamic limit}

\label{sec:tdlintro}

In this Section we explain the central statements about current operators. We do not
specify the concrete integrable model, and we consider both discrete
or continuous models (spin chains, and relativistic or
non-relativistic field theories). The formulas for the mean values
concern equilibrium situations in the thermodynamic limit, and they
were introduced in the parallel papers \cite{doyon-ghd,jacopo-ghd}.

We consider a one dimensional integrable model defined by a Hamiltonian $H$. 
We assume that the model possesses a set of charges $\{Q_\alpha\}$ which commute 
with each other and $H$ is a member of the series.
The charges are extensive operators and their operator density is
denoted as 
$q_\alpha(x)$. Here $x$ is a space coordinate, which can be
discrete (spin chains) or continuous (field theory models); the
corresponding relations are
\begin{equation}
  Q_\alpha=\sum_x q_\alpha(x),\qquad \text{or}\qquad Q_\alpha=\int dx\ q_\alpha(x).
\end{equation}
The current operators $J_\alpha(x)$ are defined by the continuity
equation for the flow of the charge densities. 
We get the definitions
\begin{equation}
  \label{Jdef1}
  i  \left[H,  q_\alpha(x)\right]=J_\alpha(x)-J_\alpha(x+1),\qquad
  \text{or}\qquad
    i  \left[H,  q_\alpha(x)\right]=-\partial_x J_\alpha(x).
\end{equation}
Let us assume that the model can be solved by the Bethe Ansatz
(the Bethe wave function will be discussed below). Then the current
mean values take a rather simple form, which seems general, it does not depend on the
particular details of the model.

For simplicity let us assume
that the finite volume eigenstates can be described by one set of so-called Bethe rapidities. More complicated
cases with multiple types of rapidities (including the string solutions, or models solvable by the
nested Bethe Ansatz) can be treated similarly. For the finite set of Bethe rapidities we will use the
notation $\lan$, whereas in the thermodynamic limit we describe them by the root density functions
$\rho(\lambda)$.

It is known that the charges act additively on the
eigenstates. In the TDL the mean values of the charge densities read
 \begin{equation}
   \begin{split}
       \vevv{q_\alpha(x,t)}&=\int d\lambda\ h_\alpha(\lambda)   \rho(\lambda),
   \end{split}
 \end{equation}
 where $h_\alpha(\lambda)$ is the one-particle charge eigenvalue function. For the mean values of
 the current operators in an equilibrium state the following formula was given in \cite{doyon-ghd,jacopo-ghd}:
  \begin{equation}
   \label{Jconject}
    \vevv{J_\alpha(x,t)}=\int d\lambda\ \veff(\lambda) h_\alpha(\lambda)   \rho(\lambda).
 \end{equation}
Here $\veff(\lambda)$ is an effective velocity, which describes the propagation of a wave packet in
the background of the other particles present in the system. It is a generalization of the group
velocity from free models, and it is given by
\begin{equation}
   \label{veffregi}
  \veff(\lambda)=\frac{de^{\text{DR}}}{dp^{\text{DR}}},
\end{equation}
where $e^{\text{DR}}$ and $p^{\text{DR}}$ are the so-called dressed energy and dressed
momentum. They are defined as the energy and momentum increase as we add one more particle with
rapidity $\lambda$ into the sea of particles in the equilibrium state. Precise formulas will be
given later.

Equation \eqref{Jconject} seems to be rather general: it is expected that it holds for every
integrable model with particle conservation; the only variation between the models is the particle
content, and the precise formulas for the dressed quantities. The formula is reminiscent of
classical physics: the currents flowing through a point are given by the number of particles passing
the point in a given time (which is given by their densities and the speed of the particles) multiplied by the
charges that the particles carry. However, it turns out that the formula holds even in the quantum
mechanical models, and the function $\veff(\lambda)$ involves the exact solution of the quantum
model. Furthermore, $\veff(\lambda)$ depends on the particular physical situation, because the individual effective
velocities depend on the background, which consists of  all the other particles present in the system. Thus the mean
currents describe the collective motion of particles, a true many body effect.

The statement  \eqref{Jconject}  seems to hold generally in integrable
models. 
However, as far as we know there is no general and rigorous proof of
it, which would hold for
{\it every} integrable model. The reason for this is simply that the
different types of models require various techniques for their
solutions, and there is no single universal method.
However, there are a number of results available, which support or even
prove \eqref{Jconject} with increasing rigor
and scope of validity. Let us now discuss these arguments and proofs.

It was shown in \cite{maurizio-currents}  that the statement always holds in models
equivalent to free bosons or free fermions.
In interacting cases proofs were given in various settings. The
original paper \cite{doyon-ghd} included a proof for relativistic QFT, and
this proof was later extended in \cite{dinh-long-takato-ghd}. Their method is based on the so-called
LeClair-Mussardo integral series \cite{leclair_mussardo}, which describes one-point functions in arbitrary equilibrium
ensembles \cite{sajat-LM}.
The statement was
proven for the spin-current of the Heisenberg spin chains in \cite{kluemper-spin-current}. In the
case of the Toda chain the statement was discussed in \cite{spohn-toda-proof}.
The paper \cite{takato-spohn} includes a rather general proof based on the
existence of a conserved current (a current operator which is itself a
conserved charge).

All the proofs mentioned above considered the thermodynamic limit directly. An other approach was
initiated in \cite{sajat-currents}, where the finite volume mean values were computed in the
Heisenberg spin chain. Here a new formula was found, which can be considered as the finite volume
origin of \eqref{Jconject}; this formula will be reviewed in Section \ref{sec:main} below. The main result
of \cite{sajat-currents} was re-derived in integrable QFT in 
\cite{bajnok-vona-currents}; their exact results also involves certain field theoretical correction terms
which are not present in the non-relativistic setting of \cite{sajat-currents}.

An alternative derivation for the finite volume mean values was found in \cite{sajat-longcurrents},
where a connection was found to the theory of long range deformed spin chains. Finally, in
\cite{sajat-algebraic-currents} a new algebraic construction was found for the current operators,
which enables the computation of their mean values using standard steps of Algebraic Bethe Ansatz.

Before moving to the next Sections let us also introduce the
generalized current operators $J_{\alpha,\beta}(x)$.
They describe the flow of the charge
$Q_\alpha$ under the time evolution generated by $Q_\beta$
\cite{benjamin-takato-note-ghd,sajat-currents}. They are defined
through the operator equation 
\begin{equation}
  \label{Jab}
  \begin{split}
 i  \left[Q_\beta, q_\alpha(x)\right]=J_{\alpha,\beta}(x)-J_{\alpha,\beta}(x+1),\qquad  \text{or}\qquad
      i  \left[Q_\beta, q_\alpha(x)\right]=-\partial_x J_{\alpha,\beta}(x).
  \end{split}
\end{equation}
For the mean values of generalized currents the following conjecture
holds:
 \begin{equation}
   \label{Jconjectg}
    \vevv{J_\alpha(x,t)}=\int d\lambda\ \veff^\beta(\lambda) h_\alpha(\lambda)   \rho(\lambda),
 \end{equation}
where now
\begin{equation}
   \label{veffregig}
  \veff^\beta(\lambda)=\frac{dQ^\beta_{\text{dr}}}{dp_{\text{dr}}}.
\end{equation}
This is an intermediate generalization of the effective velocity under
the physical time evolution.

\section{Current mean values in finite volume}

\label{sec:main}

In this Section we review the Bethe Ansatz solution of integrable
models, and we present the main statements about the current mean
values in finite volume. 

\subsection{The Bethe Ansatz}

The Bethe Ansatz is a method invented by H. Bethe in 1931
\cite{Bethe-XXX} which gives the exact eigenstates of a large class of theories.
The method can be applied in integrable models where particle number
is conserved and where the scattering of the particles is completely
elastic and factorized \cite{Mussardo-review}. In such cases the exact
wave function is relatively simple, and its functional form does not
depend on the particular details of the model.
Below we present the Bethe Ansatz solution for
simple models, where the excitations do not have an inner degree of
freedom. Extension to the multi-component cases is more complicated
and it involves the so-called nested Bethe Ansatz, which we do not
treat here.

Let us fix a reference state, which can be a
ferromagnetic state in spin chains, or the Fock vacuum in field
theories. We consider excitations above this vacuum. In spin chains
the excitations are elementary spin waves, whereas in field theories
they are the fundamental particles of the given model. In the sector with $N$ excitations let $x_1,x_2,\dots,x_N$
denote the position of the particles. We require that the coordinates
have a strict ordering:
\begin{equation}
  x_1< x_2<\dots < x_N,\qquad\text{ or }\qquad  x_1\le x_2\le \dots \le x_N.
\end{equation}
Coinciding coordinates are forbidden in the spin-1/2 chains, where
there is at most one excitation at a given site. However, coinciding
coordinates are sometimes allowed in other models.

In field theories the wave functions can be extended to an arbitrary
ordering using the bosonic or fermionic symmetry of the wave function,
but for spin chains it is not necessary to discuss alternative
orderings.

The general form of the Bethe Ansatz wave function (with the given
ordering for the coordinates) is \cite{Bethe-XXX,Korepin-Book}
\begin{equation}
  \label{bethewave}
  \Psi(x_1,\dots,x_N)=
\sum_{\mathcal{P} \in S_N} \left[ \prod_{j=1}^N e^{ix_jp_{\mathcal{P}_j}}
    \mathop{\prod_{j<k}}_{\mathcal{P}_j>\mathcal{P}_k} S(p_j,p_k)  \right].
\end{equation}
Here $p_j$ are the one-particle momenta, and $S(p_j,p_k)$ is the
two-body scattering phase, and the sum runs over all permutations of
the set of momenta.

The formula can be interpreted as follows.
The fundamental particles move freely as long as any two of them are well separated from each other.
In the  sum over permutations  each term corresponds to
a specific spatial ordering of the particles. For each permutation $\mathcal{P}\in S_N$ the number
$\mathcal{P}_j$ denotes the final position of the particle with index
$j$.
In the wave function there is a scattering factor $S(p_j,p_k)$ for
each two-body exchange.
This factor is extracted from the exact solution of the two-body problem.
The wave function is sometimes called {\it two-body irreducible}, and it reflects the factorized
scattering which is a general property of integrable models.

In many models it is possible to introduce the rapidity
parametrization, such that the scattering phase will depend only on
rapidity differences. Therefore let $\lambda$ stand for the rapidity
parameter, characterizing the momentum as $p(\lambda)$, and we assume
that the scattering phase is expressed as
\begin{equation}
  S(p_j,p_k)=S(p(\lambda_j),p(\lambda_k))=S(\lambda_j-\lambda_k)=e^{i\delta(\lambda_j-\lambda_k)}.
\end{equation}

Putting the model into a finite volume and imposing periodic boundary
conditions we obtain the Bethe equations
\begin{equation}
  \label{Betheeq}
  e^{ip(\lambda_j)}\prod_{k\ne j} S(\lambda_j-\lambda_k)=1,\qquad j=1,\dots,N.
\end{equation}
These equations serve 
as magnetization conditions for the momenta of the particles.
Putting them into logarithmic form we get
\begin{equation}
  \label{Bethelog}
  p(\lambda_j)+\sum_{k\ne j} \delta(\lambda_j-\lambda_k)=2\pi I_j,\qquad I_j\in \egesz.
\end{equation}
Here $I_j$ are the momentum quantum numbers. 

The total energy of the eigenstates is given by
\begin{equation}
  \label{Edef}
  E=\sum_{j=1}^N e(\lambda_j),
\end{equation}
where $e(\lambda)$ is the single particle energy.

The total charge eigenvalues are
\begin{equation}
  \label{Qeig}
  Q_\alpha\ket{\lan}=\Lambda_\alpha\ket{\lan}, \qquad
  \Lambda_\alpha=\sum_{j=1}^N h_\alpha(\lambda_j),
\end{equation}
where $h_\alpha(\lambda)$ is the single particle eigenvalue of the
charge $Q_\alpha$.

The additivity of these mean values follows from the extensivity and locality of the
charges, and from the mutual commutativity \cite{Mussardo-review,Korepin-Book}.

\subsection{Mean values of the currents}

It is our goal to compute the mean currents in the finite volume Bethe
states. This is a rather general problem, where the precise details of
the model do not play any role. Instead, the key ingredients are the
Bethe Ansatz wave function \eqref{bethewave} and the Bethe equations.

First we introduce the so-called Gaudin matrix, which is given by
the Jacobian of the Bethe equations
\begin{equation}
  \label{Gdef}
  G_{jk}=\frac{\partial (2\pi I_j)}{\partial \lambda_k}.
\end{equation}
Here we treated the $I_j$ quantum numbers as functions of the rapidities. For further use we spell out the matrix 
elements:
\begin{equation}
  \label{Greszletes}
  G_{jk}=\delta_{jk}\left[
p'(\lambda_j)L+\sum_{l} \varphi(\lambda_j-\lambda_l)
  \right]-\varphi(\lambda_j-\lambda_k),
\end{equation}
where we introduced the scattering kernel
\begin{equation}
  \label{varphidef}
  \varphi(\lambda)=\delta'(\lambda).
\end{equation}
The Bethe states are distributed uniformly in the space of momentum
quantum numbers $I_j$, thus the Gaudin determinant $\det G$ describes
the density of states in rapidity space. Furthermore, in many models $\det G$ is proportional to the norm of the
wave function \eqref{bethewave} \cite{Gaudin-LL-norms,XXZ-gaudin-norms,korepin-norms}.

For the mean values of the currents the following exact result was found in \cite{sajat-currents}:
\begin{equation}
  \label{main}
  \bra{\lan}J_\alpha(x)  \ket{\lan}=
{\bf e'} \cdot G^{-1} \cdot {\bf h}_\alpha.
\end{equation}
Here the quantities ${\bf e'}$
and ${\bf h}_\alpha$ are $N$-dimensional vectors with elements
\begin{equation}
  ({\bf e'})_j=\frac{\partial e(\lambda_j)}{\partial \lambda},\qquad
  ({\bf h}_{\alpha})_{j}=h_\alpha(\lambda_j),
\end{equation}
and $G^{-1}$ is the inverse of the Gaudin matrix.

For the generalized current operators the
following result was derived in \cite{sajat-currents}:
\begin{equation}
  \label{mostgeneral}
  \bra{\lan}J_{\alpha,\beta}(x)  \ket{\lan}=
{\bf h}'_\beta \cdot G^{-1} \cdot {\bf h}_\alpha.
\end{equation}
Here ${\bf h}'_\beta$ is an $N$-element vector with components
$h_\beta'(\lambda_j)$, and prime denotes
differentiation.

These formulas can be considered a direct finite volume origin of the results \eqref{Jconject}. This is explained
below. 

\subsection{Interpretation}

The formula \eqref{main}  has a simple semi-classical interpretation. Let us write it as
\begin{equation}
  \label{maskepp}
  \bra{\lan}J_\alpha(x)  \ket{\lan}=
 \frac{1}{L} \sum_{j=1}^N
 \veff(\lambda_j) h_\alpha(\lambda_j),
\end{equation}
where we defined the quantities $\veff(\lambda_j)$ as the elements of the co-vector
\begin{equation}
  \label{veffvect}
  {\bf v}_{\text{eff}}={\bf e}'\cdot G.
\end{equation}
Now we show that these quantities can be understood as effective velocities. 

First of all we note the identity
\begin{equation}
  \label{veff1}
  \veff(\lambda_j) =  \frac{L}{2\pi} \frac{\partial E}{\partial I_j},
\end{equation}
where $E$ is the total energy of the state and $I_j$ are the momentum
quantum numbers. This identity follows easily from \eqref{Edef} and \eqref{Gdef}.

Let us then discuss the structure of the formulas \eqref{veffvect}-\eqref{veff1}. In
the case of free models we have $G_{jk}=\delta_{jk}p'(\lambda_j)L$ and
$p_jL=2\pi I_j$ thus 
\begin{equation}
   \veff(\lambda_j)=\frac{e'(\lambda_j)}{p'(\lambda_j)}=\frac{de}{dp}(\lambda_j)
 \end{equation}
 is the group velocity of wave packets.
 
In the interacting case we see that the additional terms in $G$ result in corrections to the group
velocity. The resulting effective velocity can be understood in a semi-classical picture, which was
formulated in \cite{sajat-currents}. The finite volume reasoning of
\cite{sajat-currents} is analogous to the argument in the infinite
volume case given earlier in \cite{flea-gas}.  The argument goes as
follows.

Consider the
motion of $N$ particles in a periodic volume of length $L$. In the semi-classical picture we focus
on the positions of the wave packets. As the particles move around the volume, they scatter on each
other. In integrable models every multi-particle scattering event is a product of two-body
scatterings, such that the result is independent of the order of the two-body events. In our cases
each two-body scattering introduces a phase $\delta(\lambda_j-\lambda_k)$ to the wave function. It
is known that such phases result in displacements of the wave packets, which are proportional to
the scattering kernel $\varphi=\delta'$. The displacements can be understood also as time delays. In
a finite volume with a finite number of particles we intend to compute the average current of a
specific charge flowing through a selected point. In order to do this we need to focus on the long
time limit and compute the average
number of how many times each particle crosses our point of observation. This number
can be computed from the effective velocities, which describe the average propagation of the wave
packets, taking into account the displacements that accumulate along the orbits. It was shown in
\cite{sajat-currents} that a self-consistent computation for $\veff$ results exactly in the formula \eqref{veff1}.

The reader might argue that the semi-classical picture is
problematic, because the spreading of the wave packet is completely
ignored even though we are considering the large time limit in a
finite volume. This objection is actually relevant, and the problem of the diffusion of
the wave packet is already present in the derivation of
\cite{flea-gas}. Perhaps the rigorous methods of
\cite{Bajnok-classical-limit} could give a more rigorous semi-classical treatment.

In Section \ref{sec:tdl} it will be shown that in the thermodynamic
limit \eqref{veff1} turns into the conjectured formula
\eqref{veffregi}. 
Here we put forward a key idea that connects the finite and infinite
volume computation:
The correspondence between the formulas holds, because
small changes in the dressed momentum and dressed
energy can be 
traced back to small changes in the momentum quantum numbers and the
overall finite volume energy, respectively:
\begin{equation}
  \label{iden}
  \delta e^{\text{DR}}(\lambda_j)\sim \delta E,\qquad
  \delta p^{\text{DR}}(\lambda_j)\sim \delta\left(\frac{2\pi I_j}{L}\right).
\end{equation}
Thus $de^{\text{DR}}/dp^{\text{DR}}$ can be identified with $\veff$ from formula \eqref{veffvect}.

\subsection{Conjecture for more general situations}

It is tempting to formulate a general conjecture valid for a wider
range of integrable models.
Let us allow both lattice and continuum models, be it a
relativistic or non-relativistic field theory.

Let us assume that the eigenstates of the model in question can be
characterized by sets of rapidities $\{\boldsymbol
  \lambda^{(a)}_{N_a}\}_{a=1,2,\dots}$. Here the index $a$ refers to a
  {\it particle species} or {\it rapidity type}, and it is understood
  that there are $N_a$ rapidities for each type. Let us further assume
  that the eigenvalues of conserved charges can be expressed as
  \begin{equation}
    \label{Qeig2}
    Q_\alpha \ket{\{\boldsymbol  \lambda^{(a)}_{N_a}\}}=
\left[    \sum_{a} \sum_{j=1}^{N_a} h_\alpha^{(a)}(\lambda_j^{(a)})\right]
\ket{\{\boldsymbol  \lambda^{(a)}_{N_a}\}},
\end{equation}
where now the functions $h_\alpha^{(a)}(\lambda)$ describe the
one-particle eigenvalues for the particle type $a$.

Let us further assume that in finite volume the allowed sets of rapidities are given
by the Bethe equations of the form
\begin{equation}
  \label{Ba2}
 p^{(a)}(\lambda_j^{(a)})L+\sum_{b}\sum_{k=1}^{N_b}
  \delta^{(a,b)}(\lambda_j^{(a)}-\lambda_k^{(b)})
= 2\pi  I^{(a)}_j+\dots, \qquad   I^{(a)}_j\in \egesz.
\end{equation}
Here the dots stand for correction terms, that decay exponentially
with the volume, assuming that the particle number $N$ is fixed. 

Then our most general conjecture for the mean values of the currents
and generalized currents is
\begin{equation}
  \label{mostmostgeneral}
  \bra{\{\boldsymbol  \lambda^{(a)}_{N_a}\} }J_{\alpha,\beta}(x)  \ket{ \{\boldsymbol  \lambda^{(a)}_{N_a}\} }=
{\bf h}'_\beta \cdot G^{-1} \cdot {\bf h}_\alpha+\dots,
\end{equation}
where now ${\bf h}_\alpha$ is a vector of length $N=\sum_a N_a$ made of
the one-particle eigenvalues (similarly for ${\bf h}'_\beta$), and the
elements of the generalized Gaudin matrix of size $N\times N$ are
\begin{equation}
  G_{(a,j),(b,k)}=\delta_{ab}\delta_{jk}\left[
    \frac{\partial p^{(a)} (\lambda^{(a)}_j)}{\partial \lambda}L+\sum_c \sum_l \varphi^{(a,c)}_{jl}\right]-
   \varphi^{(a,b)}_{jk},
 \end{equation}
 where
 \begin{equation}
   \varphi^{(a,c)}_{jl}=\varphi^{(a,c)}(\lambda^{(a)}_j-\lambda^{(c)}_l),\qquad
   \varphi^{(a,c)}=\frac{\partial \delta^{(a,c)}(\lambda)}{\partial\lambda}.
 \end{equation}
Again, the dots in \eqref{mostmostgeneral} stand for exponentially
decaying correction terms.
 
There are multiple types of theories, where such a treatment is
necessary. Examples include:

\begin{itemize}
\item {\bf Multi-component models}. There is a large class of models
  where the fundamental excitations have an inner degree of
  freedom. These models can be solved by the {\it nested Bethe
    Ansatz}, where the different 
sets of rapidities correspond to the different {\it nesting levels},
describing the orientation in the internal space of the excitations. An example for such cases was considered in
\cite{sajat-longcurrents}, where the current mean values of the $SU(3)$-symmetric
fundamental spin chain were considered, and a result of the form \eqref{mostmostgeneral} was found.
\item {\bf Models with bound states.} In most of the spin chains that
  we consider the fundamental particles can form bound states, which
  are described by the so-called {\it string solutions}.
The bound states are stable particles, thus the different string configurations should be handled as different
particle types. Energy and momentum of the string is then parametrized
by the {\it string  center}, and we expect that formulas of the type \eqref{Qeig2}
should hold for the string centers. This will be treated in 
Section \ref{sec:xxx} below.
\item {\bf Relativistic QFT}. In integrable field theory there are always
  exponential corrections to the Bethe equations,
which   originate in virtual processes with
virtual particles winding around the volume. We expect that
\eqref{Qeig2} should hold in such relativistic situations, up to
exponentially small corrections. 
For models with one particle species it was argued in
\cite{bajnok-vona-currents} that this relation indeed holds.
\end{itemize}

We expect that if  \eqref{Ba2} and \eqref{Qeig2} hold exactly, then 
\eqref{mostmostgeneral} is also exact. Otherwise there are
exponentially small correction terms to the r.h.s. of \eqref{mostmostgeneral}.

\section{Thermodynamic limit -- general treatment}

\label{sec:tdl}

Now we perform the thermodynamic limit on the formula \eqref{main},
focusing on simple cases where the solutions to the Bethe equations
are given by real rapidities. The steps that we present here are
rather standard, but for the particular problem of current mean values
they were not yet published in the literature. The special case of the
Heisenberg spin chain will be treated in Section \ref{sec:xxx}.

We consider a situation where  in the thermodynamic limit the Bethe roots condense on the
real line. In such a case the states can be
described by the root densities $\rho(\lambda)$.
The normalization of this function is chosen such that in a large
volume $L$ the number of rapidities between $\lambda$ 
and $\lambda+\Delta \lambda$ is $L \rho(\lambda)\Delta\lambda$. Furthermore let us introduce the
hole density $\rho_h(\lambda)$; a hole is a rapidity which could be a solution to the Bethe equations
with the given set of particles, but which is itself not included in the set.

It follows from the Bethe equations \eqref{Bethelog} that
\begin{equation}
  \label{rhoeq1}
  \rho(\lambda)+\rho_h(\lambda)=\frac{1}{2\pi}p'(\lambda)
  +\int \frac{d\omega}{2\pi}\varphi(\lambda-\omega)\rho(\omega),
\end{equation}
The equation \eqref{rhoeq1} does not tell anything about the physical nature of the state in
question; it should be regarded simply as the thermodynamic representation of the fundamental Bethe
equation \eqref{Bethelog}.
On the other hand, the physical situation is specified by the {\it filling fractions}
\begin{equation}
  \label{fdef}
f(\lambda)=\frac{\rho(\lambda)}{\rho(\lambda)+\rho_h(\lambda)}.
\end{equation}
Using this definition the relation \eqref{rhoeq1} is rewritten as
\begin{equation}
  \label{rhot}
  \rho^t(\lambda)-\int \frac{d\omega}{2\pi}\varphi(\lambda-\omega)f(\omega) \rho^t(\omega)=
  \frac{1}{2\pi}p'(\lambda),
  \end{equation}
where we also introduced the total density
\begin{equation}
  \rho^t(\lambda)=\rho(\lambda)+\rho^h(\lambda).
\end{equation}
The relation \eqref{rhot} can be written in compact notation 
\begin{equation}
   (1-\hat\varphi \hat f) \rho^t=  \frac{1}{2\pi}p',
\end{equation}
where $\hat f$ and $\hat \varphi$ are the linear operators on functions acting with multiplication
with $f$ and convolution with $\varphi$, respectively. 

Let us now compute the action of the Gaudin matrix in the TDL. Let
${\bf v}$ be a vector of length $N$ with elements given by
$v(\lambda_j)$, where 
$v(\lambda)$ is a continuous function. We intend to compute the action
\begin{equation}
  G{\bf v}.
\end{equation}
First we write the matrix $G$ as
\begin{equation}
  \label{Gdecomp}
  G=D-\Phi,
\end{equation}
where $D$ is a diagonal matrix with elements
\begin{equation}
  D_{jj}=p'(\lambda_j)L+\sum_{l} \varphi(\lambda_j-\lambda_l)
\end{equation}
and $\Phi_{jk}=\varphi(\lambda_j-\lambda_k)$. It follows from the construction that in the TDL
\begin{equation}
  D_{jj}=2\pi L \rho^t(\lambda)\times (1+\ordo(L^{-1})).
\end{equation}
Furthermore, multiplication with the matrix $\varphi$ turns into a
convolution. To be precise
\begin{equation}
  \sum_{k} \varphi_{jk}
v(\lambda_k)
\quad\to\quad L\int d\lambda\  \varphi(\lambda_j-\lambda)\rho(\lambda)v(\lambda)
= L\int d\lambda\ \varphi(\lambda_j-\lambda) f(\lambda)\rho^t(\lambda)v(\lambda).
\end{equation}
Altogether we find that the elements of the vector  
\begin{equation}
  {\bf z}=G{\bf v}
\end{equation}
are given by
\begin{equation}
  z_j=
 2\pi L z(\lambda_j) \times (1+\ordo(L^{-1}))
\end{equation}
with
\begin{equation}
  \label{dress0}
  z(\lambda)=   \rho^t(\lambda) v(\lambda)-
  \int \frac{d\lambda'}{2\pi} f(\lambda')  \rho^t(\lambda) \varphi(\lambda-\lambda')v(\lambda').
\end{equation}
It is useful to introduce the ``dressing equation'' for $\tilde v(\lambda)=\rho^t(\lambda)
v(\lambda)$, which reads
\begin{equation}
  \label{dress1}
  z(\lambda)=  \tilde v(\lambda)-
  \int \frac{d\lambda'}{2\pi} f(\lambda') \varphi(\lambda-\lambda')\tilde v(\lambda').
\end{equation}
This is structurally equivalent to \eqref{rhot}. We call the solution
of the integral equation ``dressing'' and denote
\begin{equation}
  \label{dress1b}
\tilde v(\lambda)=z^\dr(\lambda).
\end{equation}
Regarding the matrices we find the following relation:
\begin{equation}
  \label{lim1}
GD^{-1}\quad\to\quad    (1-\hat\varphi \hat f).
\end{equation}
It follows that for any two functions $a(\lambda)$, $b(\lambda)$ and
the corresponding vectors ${\bf a}$ and ${\bf b}$ the following holds:
\begin{equation}
  \begin{split}
     \label{Gitdl}
\tdl  {\bf a}G^{-1}{\bf b}
=&a\cdot \hat f (1-\hat \varphi\hat f)^{-1} \cdot b= a\cdot  (\hat f^{-1}-\hat \varphi)^{-1} \cdot b=\\
=&\int \frac{d\lambda}{2\pi} f(\lambda) a(\lambda)b^\dr(\lambda)
=\int \frac{d\lambda}{2\pi} f(\lambda) a^\dr(\lambda)b(\lambda).
 \end{split}
\end{equation}
The l.h.s. of \eqref{Gitdl} is symmetric with respect to ${\bf  a}$
and ${\bf b}$. This symmetry is respected by the third expression, thus we can act the dressing
procedure on either function.

The expression ``dressing'' goes back to the problem of
adding/removing fundamental excitations to/from a sea of
particles. The dressing operation describes the change in the charge
eigenvalues in this process, and it takes into account both the charge
carried by the added/removed particle, and the backflow generated by
this process.
Let us now discuss this connection in detail, first going back to the
finite volume situation.

We consider the addition of an extra particle with rapidity $u$ to a Bethe state with
rapidities $\lan$. The presence of the new particle will shift the rapidities of the
original particles. The new rapidities will be
\begin{equation}
  \tilde\lambda_1,\dots,\tilde\lambda_N,u.
\end{equation}
We assume that the shift is continuous
and of order $1/L$. Thus we write
\begin{equation}
  \tilde\lambda_j-\lambda_j=\frac{F(\lambda_j|u) }{L},
\end{equation}
where $F(\lambda|u)$ is the so-called shift function.  It can be computed from the 
Bethe equations for the original particles. Expanding those equations to first order in $1/L$ we get
\begin{equation}
  \begin{split}
    p'(\lambda_j) F(\lambda_j|u)+\frac{1}{L}\sum_{k\ne j}
    \varphi(\lambda_j-\lambda_k)(F(\lambda_j|u)-F(\lambda_k|u))+\delta(\lambda_j-u) &=0,
  \end{split}
\end{equation}
where we used that the momentum quantum numbers corresponding to the
original particles are not changed.
In the TDL this equation can be expressed as
\begin{equation}
  2\pi \rho_t(\lambda_j) F(\lambda_j|u)-\int d\lambda\ \varphi(\lambda_j-\mu) \rho(\mu)F(\mu|u)=-\delta(\lambda_j-u).
\end{equation}
Introducing
\begin{equation}
 2\pi \rho_t(\lambda) F(\lambda|u)= \tilde F(\lambda|u)
\end{equation}
we get
\begin{equation}
 \tilde F(\lambda|u)-\int \frac{d\mu}{2\pi} \varphi(\lambda-\mu) f(\mu) \tilde F(\mu|u)=-\frac{1}{2\pi}\delta(\lambda-u).
\end{equation}
We remind that the $\delta$-function here describes the scattering phase shift, and not a Dirac-delta.

Let us now consider the energy difference as an effect of the
addition. Calling it ``dressed energy'' we get the approximate finite
volume expression
\begin{equation}
  \edr(u)=e(u)+\frac{1}{L}\sum_j e'(\lambda_j) F(\lambda_j|u).
\end{equation}
The notation $^{\text{DR}}$ is not a typo here: this dressing is different
from the one defined in \eqref{dress1b}. The connection is explained below.

In the TDL the dressed energy is expressed as
\begin{equation}
  \edr(u)=e(u)+\int \frac{d\lambda}{2\pi} e'(\lambda) f(\lambda) \tilde F(\lambda|u).
\end{equation}
The equations simplify if we consider the derivative of the dressed
energy with respect to the rapidity. For example we obtain
\begin{equation}
  \partial_u  \edr(u)=\partial_u e(u)+\int \frac{d\lambda}{2\pi} e'(\lambda) f(\lambda) \partial_u\tilde F(\lambda|u)
\end{equation}
and for the derivative of the shift function we get
\begin{equation}
\partial_u \tilde F(\lambda|u)-\int \frac{d\mu}{2\pi} \varphi(\lambda-\mu) f(\mu) \partial_u\tilde F(\mu|u)
=\frac{1}{2\pi}\varphi(\lambda-u).
\end{equation}
From this the following simple relation is found:
\begin{equation}
   \partial_u  \edr(u)-\int \frac{d\mu}{2\pi} \varphi(u-\mu) f(\mu)\partial_\mu\edr(\mu)=e'(u).
 \end{equation}
 This means that
 \begin{equation}
   \partial_u \edr(u)=(e'(u))^{\text{dr}}.
 \end{equation}
This relation connects the two different definitions for the dressing.

For the dressed momentum we obtain
\begin{equation}
   \partial_u  p^{\text{DR}}(u)-\int \frac{d\lambda}{2\pi} \varphi(u-\mu) f(\mu)\partial_\mu p^{\text{DR}}(\mu)=p'(u).
\end{equation}
Combining with \eqref{rhoeq1} we get
\begin{equation}
  \label{ppp}
   \partial_u  p^{\text{DR}}(u)=2\pi \rho_t(u).
 \end{equation}

Let us now compute the current mean values. Applying the general
formula \eqref{Gitdl} to the mean value \eqref{main} we get
\begin{equation}
  \vevv{J_\alpha}=\int \frac{d\lambda}{2\pi} f(\lambda) (e')^\dr(\lambda)h_\alpha(\lambda).
\end{equation}
Using the definition \eqref{fdef} and \eqref{ppp} this is written as
\begin{equation}
  \vevv{J_\alpha}=\int d\lambda\ \rho(\lambda)
\frac{(e')^\dr(\lambda)}{\partial_\lambda  p^{\text{DR}}(\lambda)}h_\alpha(\lambda)  
=\int d\lambda\ \rho(\lambda)
\frac{\partial_\lambda \edr(\lambda)}{\partial_\lambda  p^{\text{DR}}(\lambda)}h_\alpha(\lambda).
\end{equation}
Making the identification \eqref{veffregi} we obtain the original conjecture \eqref{Jconject}.

\section{Current mean values in the Heisenberg spin chain}

\label{sec:xxx}

In this Section we treat the Heisenberg spin chains: we discuss the
charges and currents of the model. The Section includes new results
for the currents which were not yet presented in the literature: the
discussion of the finite volume formulas in terms of the string
solutions is new, together with the corresponding thermodynamic limit.

The so-called XXZ Heisenberg chain is defined by the Hamiltonian
\begin{equation}
  \label{HXXZ0}
  H=\sum_j \sigma^x_j\sigma^x_{j+1}+ \sigma^y_j\sigma^y_{j+1}+\Delta (\sigma^z_j\sigma^z_{j+1}-1).
\end{equation}
Here $\Delta$ is the so-called anisotropy parameter. The
special point $\Delta=1$ describes the $SU(2)$-invariant XXX model, originally solved by Bethe
\cite{Bethe-XXX}. In this review we focus on the XXX case, where the fundamental particles can form bound states of
arbitrary size. The XXZ model with $\Delta>1$ can be treated analogously, but
the so-called massless case with $|\Delta|<1$ requires a separate treatment due to the delicate bound state structure of
that model \cite{Takahashi-Book}.

The Heisenberg chains are solved by the Bethe Ansatz. The reference state is chosen as a
ferromagnetic state, for example a state with all spins up. Then the fundamental excitations are the
propagating waves formed by the down spins. The exact finite volume eigenstates take the form
\eqref{bethewave}; the characteristic functions in the XXX model are  
\begin{equation}
  \label{XXZfunct}
  \begin{split}
  e^{i p(\lambda)}&= \frac{\lambda-i/2}{\lambda+i/2}\qquad
S(\lambda)=e^{i\delta(\lambda)}=\frac{\lambda+i}{\lambda-i}.
\end{split}
\end{equation}

The charge and current operators in these spin chains can be obtained from the Quantum Inverse
Scattering Approach. This is treated in Section \ref{sec:algebraic}. We just put forward that in a
particular normalization the
one-particle eigenvalues of the charges are
\begin{align}
  \label{halpha0}
    h_2(\lambda) &= - \frac{1}{\lambda^2 + 1/4} \\
    h_\alpha(\lambda) &= \partial_x^{\alpha-2}\ h_2(\lambda-x) \Big\rvert_{x=0} \hspace{1 cm} \alpha > 2.
\end{align}
Furthermore
\begin{equation}
  e(\lambda)=2h_2(\lambda).
\end{equation}
Then mean values of the currents are given by formula
\eqref{main}.
Here we do not discuss the proof of the finite volume formula, which is later presented in
 Sections \ref{sec:fftcsa2} and \ref{sec:algebraic}. 
Instead,
we perform a closer analysis of the formula, with the goal of taking
the proper thermodynamic limit. This consists of two
steps. The first step is to take into account the bound state structure of the XXX chain: it is
known that the model allows for string solutions with arbitrary length, that should be treated as
separate particle types. We will show that the currents can be expressed in the form \eqref{mostmostgeneral}, which
was put forward on general grounds. As a second step we will take the thermodynamic limit of that
formula. Thereby we derive a new compact expression for the currents.

Before turning to the actual computations let us introduce generating
functions for the charges and the currents. We will see that it is
more convenient to work with them. For the charges we introduce the
formal sum
\begin{equation}
  \label{qnudef}
  Q(\nu)=\sum_{\alpha=2}^\infty \frac{\nu^{\alpha-2}}{(\alpha-2)!} Q_\alpha.
\end{equation}
It is shown in Section \ref{sec:algebraic} that this is a very natural
definition which is compatible with the transfer matrix
construction. This generating function is also an extensive operator,
and its operator density is defined as
\begin{equation}
  q(\nu,x)=\sum_{\alpha=2}^\infty \frac{\nu^{\alpha-2}}{(\alpha-2)!} q_\alpha(x).
\end{equation}
Regarding the currents we focus on the generalized operators
$J_{\alpha,\beta}$ defined in \eqref{Jab}. It is natural to introduce
a generating function with two spectral parameters:
\begin{equation}
  \label{Jsum}
 J(\mu,\nu,x)=\sum_{\alpha=2}^\infty \sum_{\beta=2}^\infty\frac{\mu^{\alpha-2}}{(\alpha-2)!}
  \frac{\nu^{\beta-2}}{(\beta-2)!}  J_{\alpha,\beta}(x).
\end{equation}
This two-parameter family of operators satisfies the generalized
continuity relation
\begin{equation}
  \label{Jmunudef}
  i  \left[ Q(\nu),q(\mu,x)\right]= J(\mu,\nu,x)- J(\mu,\nu,x+1).
\end{equation}

\subsection{String solutions in finite volume}

The solutions of the Bethe equations organize themselves into the so-called
strings in the complex plane. A string describes a bound state of the fundamental spin waves, where the rapidities
within the string follow a well defined pattern, dictated by the pole of the $S$-matrix. In the XXX case this means that
within each string the rapidities follow each other with a difference of $i$ at each step.

In the presence of multiple strings  each rapidity is identified as the $r$-th element
of the $\varrho$-th string of length $j$: 
\begin{equation}
    \lambda_{j,\varrho}^r = \lambda_{j,\varrho} + \frac{i}{2} (j + 1 - 2r) + \delta_{j,\varrho}^r.
\end{equation}
Here $\lambda_{j,\varrho}\in\valos$ is the string center and $\delta_{j,\varrho}^r$ is the so-called
string deviation which vanishes exponentially with the system size. Different string lengths
correspond to different types of excitations, interpreted as bound states of elementary particles. 

Considering the string structure one can reformulate the Bethe equations and the Gaudin matrix as
follows \cite{Takahashi-Book}. Reduced equations for the string centers (also called Bethe-Takahashi
equations) can be obtained by taking products of Bethe equations for the constituents within the strings.
Thus we get
\begin{equation}
    e^{ip_j(\lambda_{j,\varrho})L} \prod_{\substack{ (k, \sigma) \\ \neq (j,\varrho) }} S_{j,k}(\lambda_{j,\varrho}-\lambda_{j,\sigma}) = 1,
\end{equation}
or in the logarithmic form
\begin{equation}
    L \ln \Big( ip_j(\lambda_{j,\varrho}) \Big) + \sum_{\substack{(k,\sigma)\\\neq (j,\varrho)}} \ln S_{j,k}(\lambda_{j,\varrho}-\lambda_{k,\sigma}) = 2\pi i I_{j,\varrho}.
\end{equation}
These lead to the reduced Gaudin matrix:
\begin{gather}
    G_{(j,\varrho),(k,\sigma)}^{\text{(r)}} = \frac{\partial(2\pi I_{j,\varrho})}{\partial \lambda_{k,\sigma}} = \\
    = \delta_{(j,\varrho),(k,\sigma)} \Bigg( Lp'_j(\lambda_{j,\varrho}) + \sum_{\substack{(l,\tau)\\ \neq (j,\varrho)}} \varphi_{j,l}(\lambda_{j,\varrho} - \lambda_{l,\tau}) \Bigg) - (1-\delta_{(j,\varrho),(k,\sigma)}) \varphi_{j,k}(\lambda_{j,\varrho}-\lambda_{k,\sigma}).
\end{gather}
Let us define a family of functions
\begin{equation}
    a_j(\lambda) = \frac{4j}{4\lambda^2 + j^2},\qquad j=1,2,\dots \label{eq:a_functions}
\end{equation}
Then the derivatives that enter the Gaudin matrix are given by
\begin{equation}
    p'_j(\lambda) = a_j(\lambda)
\end{equation}
and
\begin{align}
    \varphi_{j,k}(\lambda) = -
    \begin{cases}
    a_{\abs{j-k}}(\lambda) + &2a_{\abs{j-k}+2}(\lambda) + ... + 2a_{j+k-2}(\lambda) + a_{j+k}(\lambda) \hspace{1cm} j\neq k \\
    &2a_{\abs{j-k}+2}(\lambda) + ... + 2a_{j+k-2}(\lambda) + a_{j+k}(\lambda)  \hspace{1 cm} j=k.
    \end{cases}
\end{align}
These formulas can be obtained for example by a summation of the elements of the
original Gaudin matrix, corresponding to the constituents of each string:
\begin{equation}
    G_{(j,\varrho),(k,\sigma)}^{\text{(r)}} = \sum_{r=1}^j \sum_{s=1}^k G_{(j,\varrho,r), (k,\sigma,s)}.
\end{equation}
Each block of the original matrix is represented in the reduced matrix as the sum of its elements.

It was shown in \cite{kirillov-korepin-norms-strings} that the determinant of the reduced matrix
describes the norm of the Bethe state with strings. Below we show that
this matrix enters the formulas for the current mean values as well.

\subsection{Mean values of charges and currents with strings}

Here we compute the eigenvalues of charges and the mean values of currents using the string hypothesis. We focus on the 
generating functions \eqref{qnudef} and \eqref{Jsum} for the charges and currents, respectively. 

For the charge generating function $Q(\nu)$ the single particle eigenvalues are
\begin{equation}
h(\lambda|\nu) = -a_1(\lambda-\nu),
\end{equation}
where the function $a_1(u)$ is given by \eqref{eq:a_functions}.
This result can be computed from the QISM formalism, and the simple shift in $\nu$ is compatible with \eqref{halpha0}. 

The charge eigenvalue of a $j$-string with center $\lambda_{j,\varrho}$ can be computed using a telescopic sum, and we
obtain  
\begin{equation}
    h_{j} (\lambda_{j,\varrho}|\nu) = \sum_{r=1}^j h(\lambda_{j,\varrho}^r|\nu) = -a_j(\lambda_{j,\varrho}-\nu).
    \label{eq:string_charges}
\end{equation}

The current mean values can also be calculated with respect to the string centers up to exponential corrections:
\begin{equation}
    \expval{J(\mu,\nu)}{\{\lambda_{j,\varrho}\}_{j,\varrho}} = \sum_{(j,\varrho)} \sum_{(k,\sigma)} h_{j}'(\lambda_{j\varrho}|\nu) G_{(j,\varrho),(k,\sigma)}^{\text{(r)}^{-1}} h_{k}(\lambda_{k,\sigma}|\mu) + \mathcal{O}(\delta).
    \label{eq:xxx_finite_current_formula}
\end{equation}
This can be proven starting from the finite volume formula
\eqref{main} {by finding a relation between the reduced and the ordinary inverse Gaudin matrices.} Now we sketch the key steps of the proof, which is presented
with all details in Appendix \ref{sec:reduced-gaudin}.

For a complex invertible square matrix $A$ its minor corresponding to the $j$-th row and $k$-th column is defined as
\begin{equation}
    M_{jk} = \det \Big( (A_{lm})_{l\neq j, m\neq k} \Big),
\end{equation}
that is, the determinant of the matrix missing the given row and column. The corresponding co-factor is
\begin{equation}
    C_{jk} = (-1)^{j+k} M_{jk}.
\end{equation}
Using the co-factor matrix one can express the inverse of $A$ as
\begin{equation}
    A^{-1} = \frac{1}{\det A} \cdot C^T.
\end{equation}
Considering the string structure the Gaudin matrix has the explicit form
\begin{align}
    G_{(j,\varrho,r),(k,\sigma,s)} = \delta_{(j,\varrho,r), (k,\sigma,s)} &\Bigg( \frac{4L}{4\lambda_{j,\varrho}^r + 1} - \sum_{\substack{(l,\tau,t)\\\neq (j,\varrho,r)}} \frac{2}{(\lambda_{j,\varrho}^r - \lambda_{l,\tau}^t)^2+1} \Bigg) + \nonumber \\
    + &(1-\delta_{(j,\varrho,r), (k,\sigma,s)}) \cdot \frac{2}{(\lambda_{j,\varrho}^r - \lambda_{k,\sigma}^s)^2+1}.
    \label{eq:gaudin_strings}
\end{align}
Since adjacent rapidities in a string differ by a complex unit up to
exponentially vanishing corrections, the following divergent expressions appear in the matrix:
\begin{equation}
    K_{j,\varrho}^{r,r+1} := -\frac{2}{( \lambda_{j,\varrho}^{r+1} - \lambda_{j,\varrho}^r )^2 + 1}.
    \label{eq:divergences}
\end{equation}
Our goal is to find the leading order behaviour of the determinant and
co-factors in terms of these expressions. They are present in, right
above and right below the diagonal. 

To calculate the Gaudin determinant the key step is to add certain
rows and columns, in order to reduce the number
of divergent elements in the matrix. In fact, it is possible
to obtain a form when they are only present in the diagonal.  
This treatment is also applicable to the co-factors with a few
modifications and additional row and column exchanges. The latter
introduces nontrivial signs but in the end we have all divergent
elements in the diagonal again. 

After these simplifications the divergent expressions can be factored
out from the diagonal, and we can always neglect the lower order terms in
the thermodynamic limit. The remaining task is to
relate the rest to the reduced matrix.
This is possible because our
elimination process for the divergent expressions earlier had the
side-effect of summing up all elements in every block. One can show
that the remaining factor is exactly the determinant of these block
sums, that is, the determinant of the reduced matrix. The same is true
for the co-factors as well but in that case the signs must also be
matched which requires a careful analysis. 

Nevertheless, the following results can be obtained:
\begin{align}
    \det G &= \Bigg( \prod_{(j,\varrho} \prod_{r=1}^{j-1} K_{j,\varrho}^{r,r+1} \Bigg) \cdot \det G^{(\text{r})} \cdot \Big( 1 + \mathcal{O}(1/K) \Big) \\
     C_{(j,\varrho,r),(k,\sigma,s)} &= \Bigg( \prod_{(j,\varrho)} \prod_{r=1}^{j-1} K_{j,\varrho}^{r,r+1} \Bigg) \cdot C_{(j,\varrho),(k,\sigma)}^{(\text{r})} \cdot \Big( 1 + \mathcal{O}(1/K) \Big) \hspace{1 cm} \forall r,s
\end{align}
where $\mathcal{O}(1/K)$ denotes terms vanishing exponentially. For the inverse we accordingly have
\begin{equation}
    G_{(j,\varrho,r),(k,\sigma,s)}^{-1} = G_{(j,\varrho),(k,\sigma)}^{(\text{r})^{-1}} \cdot \Big( 1 + \mathcal{O}(1/K) \Big) \hspace{1 cm} \forall r,s.
\end{equation}
Interestingly, to leading order all elements are identical in a given
block of the inverse matrix. From here the formula \eqref{eq:xxx_finite_current_formula} follows
immediately. The details of this computation are presented in Appendix \ref{sec:reduced-gaudin}.

\subsection{Thermodynamic limit: XXX chain}

In the thermodynamic limit the string centers become dense for all string types, and they will be
described by density functions $\rho_j(\lambda)$ for particles and $\rho_j^h(\lambda)$ for
holes. The Bethe equations are transformed into 
\begin{equation}
    \rho_j (\lambda) + \rho_j^h(\lambda) = \frac{1}{2\pi} p_j'(\lambda) + \sum_{k=1}^\infty \int_{-\infty}^\infty \frac{d\omega}{2\pi} \varphi_{j,k}(\lambda - \omega) \rho_k(\omega).
\end{equation}
Once again we define the total root densities $\rho_j^t(\lambda)$ and the filling fractions
$f_j(\lambda)$ so that we have the alternative form of the equations 
\begin{equation}
    \rho_j^t(\lambda) - \sum_{k=1}^\infty \int_{-\infty}^\infty \frac{d\omega}{2\pi} \varphi_{j,k}(\lambda-\omega) f_k(\omega) \rho_j^t(\omega) = \frac{1}{2\pi} p_j'(\lambda).
    \label{eq:xxx_tdl_bethe_eq}
\end{equation}
Using the previous compact notation now we should write
\begin{equation}
    \big(1 - \underline{\underline{\hat{\varphi}}} \underline{\hat{f}} \big) \underline{\rho^t} = \frac{1}{2\pi} \underline{p}',
\end{equation}
where the infinite number of vector and matrix components are labeled by the string type index.

Let us examine the effect of the reduced Gaudin matrix in a similar fashion as in the real rapidity case. Consider it acting on a vector of some spectral function $v_{k,\sigma}=v_k(\lambda_{k,\sigma})$:
\begin{equation}
    z_{j,\varrho} \coloneqq \sum_{(k,\sigma)} G_{(j,\varrho),(k,\sigma)}^{(\text{r})} v_{k,\sigma}.
\end{equation}
The main difference brought by the strings is that now our linear operator in the thermodynamic limit acts on a family of functions $\{v_j\}_j$ labeled by the string index:

\begin{equation}
    z_j(\lambda) = 2\pi L \Bigg( \rho_j^t(\lambda) v_j(\lambda) - \sum_{k=1}^\infty \int_{-\infty}^\infty \frac{d\omega}{2\pi} \varphi_{j,k}(\lambda-\omega) f_k(\omega) \rho_k^t(\omega) v_k(\omega) \Bigg).
\end{equation}
Using which we once again have a dressing equation for the functions $x_j=\frac{z_j}{2\pi L}$ and $x_j^{\text{dr}}=\rho_j^t v_j$:
\begin{equation}
    x_j(\lambda) = x_j^{\text{dr}}(\lambda) - \sum_{k=1}^\infty \int_{-\infty}^\infty \frac{d\omega}{2\pi} \varphi_{j,k}(\lambda-\omega) f_k(\omega) x_j^{\text{dr}}(\omega).
    \label{eq:xxx_dressing_equation}
\end{equation}
In our compact notation this reads:
\begin{equation}
    \underline{x} = \big(1 - \underline{\underline{\hat{\varphi}}} \hat{f} \big) \underline{x}^{\text{dr}} \eqqcolon \big(1 - \underline{\underline{\hat{T}}}\big) \underline{x}^\text{dr}.
    \label{eq:xxx_dressing_equation_compact}
\end{equation}
According to these results for any two spectral functions $a_j(\lambda)$ and $b_j(\lambda)$ we have the thermodynamic limit
\begin{equation}
    \lim_{\text{TDL}} \sum_{(j,\varrho)} \sum_{(k,\sigma)} a_j(\lambda_{j,\varrho}) G_{(j,\varrho),(k,\sigma)}^{\text{(r)}-1} b_k(\lambda_{k,\sigma}) = \sum_{j=1}^\infty \int_{-\infty}^\infty \frac{d\lambda}{2\pi} f_j(\lambda) a_j(\lambda) b_j^{\text{dr}}(\lambda),
    \label{eq:xxx_gaudin_tdl}
\end{equation}
or using our compact notation
\begin{equation}
    = \underline{a} \hat{f} (1- \underline{\underline{\varphi}}\hat{f})^{-1} \underline{b},
\end{equation}
where the dot product of the two vectors also includes integration. As in the case of real rapidities, the expression is symmetric under the exchange of the spectral functions $a$ and $b$.

As a result of the string structure, all of our formulas contain an
infinite sum over the string indices. However, it is known that such
equations can be partially disentangled, leading to simplified
formulas \cite{Takahashi-Book}. For the sake of completeness, we give
all details of the decoupling procedure in Appendix
\ref{sec:decoupling}, and below we present the main results.

First let us write down the decoupled form of the integral equation
for the root densities. This reads
\begin{equation}
    \rho_j^t = \frac{1}{2\pi} \delta_{j,1} s + s * (\rho_{j-1}^h + \rho_{j+1}^h),
    \label{eq:decoupled_bethe_eqs}
\end{equation}
where the function $s$ is
\begin{equation}
    s(\lambda ) = \frac{\pi}{\cosh{\pi\lambda}}
\end{equation}
and $*$ denotes the convolution
\begin{equation}
    \Big( f*g \Big)(\lambda) = \int_{-\infty}^\infty \frac{d\omega}{2\pi} f(\omega) g(\lambda-\omega).
\end{equation}
Also $\rho_0^h = 0$ was defined. The term $\frac{1}{2\pi}\delta_{j,1} s$ is called the source term.

For the charge mean values we have
\begin{equation}
    \frac{ \expval{Q(\nu)}}{L}
    = \sum_{j=1}^\infty \int_{-\infty}^\infty d\lambda\ \rho_j(\lambda) h_{j}(\lambda|\nu) = \Big[ s*(2\pi \rho_1^h-a_1) \Big](\nu).
    \label{eq:xxx_tdl_charge_meanvalue}
\end{equation}
This relation was first derived in \cite{JS-CGGE}.

Let us now consider the dressing equations \eqref{eq:xxx_dressing_equation}. These can also be disentangled yielding 
\begin{align}
    h_{j}^\text{dr} (\lambda|\nu) = - \frac{1}{2\pi} \delta_{j,1}  s(\lambda-\nu)  &+ \nonumber \\
    + \int_{-\infty}^\infty \frac{d\omega}{2\pi}\ s(\lambda&-\omega) \Big( \eta_{j-1}(\omega)    h_{j-1}^\text{dr}(\omega|\nu)
    + \eta_{j+1}(\omega) h_{j+1}^\text{dr}(\omega|\nu) \Big),
    \label{eq:decoupled_dresing_equations-altalanos}
\end{align}
where $\eta_j=\rho_j^h/\rho_j^t$. As a result the currents become
\begin{gather}
    \expval{J(\mu,\nu)} = \sum_{j=1}^\infty \int_{-\infty}^\infty \frac{d\lambda}{2\pi}\ f_j(\lambda) h_{j}'(\lambda|\nu) h_{j}^\text{dr}(\lambda|\mu) =\\
    = - \partial_\nu \Bigg[ \int_{-\infty}^\infty \frac{d\omega}{2\pi}\ s (\nu-\omega) \eta_1 (\omega) h_{1}^\text{dr}(\omega|\mu) + \frac{1}{2\pi} \Big( a_1 * s \Big)(\nu-\mu) \Bigg].
    \label{eq:xxx_tdl_current_formula}
\end{gather}

Analogous formulas in the XXZ spin chain were already given in \cite{sajat-QA-GGE-hosszu-cikk}, in the context of
factorized correlation functions. The present formula \eqref{eq:xxx_tdl_current_formula} is equivalent (apart from
normalization) to the results of 
 \cite{sajat-QA-GGE-hosszu-cikk} for the quantity $\Omega$ presented in Section 5 of that work. At the time of the
 publication of \cite{sajat-QA-GGE-hosszu-cikk} it was not clear that these objects describe current mean values. The
 explanation for the coincidence was only found later in \cite{sajat-algebraic-currents}, and it is presented below in
 Sections \ref{sec:algebraic} and \ref{sec:corr}.

\section{Proof using form factor expansion}

\label{sec:fftcsa2}

Here we present the first proof of the main statement \eqref{main} which appeared in  \cite{sajat-currents}.
Here the idea was to focus on the off-diagonal matrix elements of the charge and current operators,
and to approach the mean values using a careful diagonal limit. 
This is possible because the form factors (matrix elements of local operators) satisfy very special
relations in integrable
models, and the mean values are very closely related to the off-diagonal elements. 
Furthermore, the continuity
relations  yield information about these off-diagonal elements, and
this can be used to construct the mean values. 

In any finite volume and for any two Bethe states with the same
particle numbers the continuity equation gives
\begin{equation}
  \label{QJ}
  \begin{split}
 &  i \left(\sum_{j=1}^N e(\lambda_j)-e(\mu_j)\right)
   \bra{\lan}Q_\alpha(x)\ket{\mun}=
 \left(1-\prod_{j=1}^N e^{i(p(\mu_j)-p(\lambda_j))}\right) \bra{\lan}J_\alpha(x)\ket{\mun}.
  \end{split}
 \end{equation}
This relation only holds for the on-shell states of a finite volume theory, where the roots
satisfy the Bethe equations. However, the relation can be extended to the infinite volume situation,
when there is no restriction on the Bethe roots.

To this order let us consider the {\it form factors} $F^\ordo(\lan|\mun)$ of an arbitrary local
operator $\ordo$, which are defined as the matrix elements between infinite volume scattering
states. It is known that the form factors are meromorphic functions which satisfy special
properties \cite{Korepin-Book}. Furthermore, the relation between the normalized off-diagonal matrix elements for
generic states $\ket{\lan}$ and $\ket{\mun}$ is 
\begin{equation}
  \label{fftcsa1rel}
  \bra{\lan}\ordo\ket{\mun}=\frac{F^\ordo(\lan|\mun)}{\sqrt{\det_N G(\lan)
      \det_N G(\mun)}},
\end{equation}
The two Gaudin determinants in the denominator stem from the norm of the
Bethe Ansatz wave functions. This relation was introduced for integrable QFT in \cite{fftcsa1}
(where it suffers additional correction terms), and for non-relativistic models it is known to hold
generally \cite{Korepin-Book}.

Let us define the
so-called symmetric evaluation of the (infinite volume) diagonal form factors of any operator as
\begin{equation}
  \begin{split}
    &  F^{\ordo}_s(\lan)=
    \lim_{\eps\to 0}
 F^\ordo(\lambda_1+\eps,\dots,\lambda_N+\eps|\lambda_N,\dots,\lambda_1).
  \end{split}  
\end{equation}

Using \eqref{fftcsa1rel} we can extend relation \eqref{QJ}
also to the form factors, where there is no restriction on the Bethe roots.
 Introducing the short-hand notations
\begin{equation}
  \begin{split}
      \FQa(\lan)&\equiv F_{s}^{Q_\alpha(0)}(\lan)\\
  \FJa(\lan)&\equiv F_{s}^{J_\alpha(0)}(\lan)
    \end{split}
\end{equation}
we get from  \eqref{QJ}
\begin{equation}
  \label{QJsym}
  \left(\sum_{j=1}^N e'(\lambda_j)\right)
  \FQa(\{\lambda\}_N)=
\left(\sum_{j=1}^N p'(\lambda_j)\right) 
\FJa(\{\lambda\}_N).
\end{equation}
The finite volume mean values can be computed from special limiting values of these objects.

It is also useful to define the functions
$\rho_N(\lambda_1,\dots,\lambda_N)$ as the $N\times N$ Gaudin determinants
evaluated at the set of rapidities $\{\lambda_1,\dots,\lambda_N\}$. In
the notations we suppress the index $N$ and write simply
\begin{equation}
\rho(\lan)=\det G(\lan).
\end{equation}
We remind that the Gaudin determinants describe the norms of Bethe
wave functions for eigenstates, ie. for sets of rapidities satisfying
the Bethe equations. On the other hand, the functions
$\rho(\lan)$ are defined for arbitrary sets of rapidities.

The finite volume mean values of local operators can be computed
through the expansion
\begin{equation}
  \label{fftcsa2}
  \bra{\lan}\ordo(0)\ket{\lan}=
\frac{\sum\limits_{\{\lambda^+\}\cup\{\lambda^-\}}
  F^{\ordo}_s(\{\lambda^+\}) \rho(\{\lambda^-\}) }
{\rho(\lan)},
\end{equation}
where the summation runs over all partitionings of the set of the
rapidities into $\{\lambda^+\}\cup\{\lambda^-\}$.
The partitionings include those cases when either
subset is the empty set, and in these cases it is understood that $\rho(\emptyset)=1$ and
 $F^{\ordo}_s(\emptyset)=\vev{\ordo}$ is the v.e.v. The relation \eqref{fftcsa2}
is exact when the Bethe Ansatz wave functions are exact eigenstates of
the model. It was first proposed in \cite{fftcsa2} in integrable QFT,
and for the integrable spin chains it was proven first in
\cite{yunfeng1} and then independently in \cite{sajat-currents}.

The key idea of \cite{sajat-currents} was to use the above expansion
theorem twice. First it can be used to extract the symmetric form
factors of the charge density operators, because the charge mean values
are known, they are given by \eqref{Qeig}. Then relation \eqref{QJ} gives the symmetric form factors
of the current operators. Finally, the expansion theorem can be used
a second time to find the explicit formula for the current mean
values.

We do not reproduce this computation here, the detailed derivations are quite
lengthy and cumbersome.
Here we just present the key statements.

The symmetric form factors of the charge and current operators can be expressed using graph
theoretical summations. The reason for this lies in the special structure of the Gaudin matrix,
which enables the use of Kirchhoff's theorem (also known as the matrix-tree theorem) for the
determinant and the inverse matrix. 

Let us introduce the following definitions.
For a graph $\Gamma$ a directed graph $\mathcal{F}$ is a directed spanning forest of $\Gamma$
if the following conditions hold:
\begin{itemize}
\item $\mathcal{F}$ includes all vertices of $\Gamma$.
\item $\mathcal{F}$ does not include any circles.
\item Each vertex has at most one incoming edge.
\end{itemize}
The nodes that do not have incoming edges are called roots. Each spanning
forest can be decomposed as a union of spanning trees, which are the
 connected components of the forest. Each spanning tree has
exactly one root. 

Using these notations the symmetric form factors of the charges are
\begin{equation}
  \label{FQs}
  \FQa(\lan)=
  \left[\sum_{j=1}^N p'(\lambda_j)\right]
  \left[\sum_{j=1}^N q_\alpha(\lambda_j)\right]
  \sum_{\mathcal{T}}  \prod_{l_{jk}\in \mathcal{T}} \varphi_{jk}.
\end{equation}
Here the summation runs over all directed spanning forests of the complete graph with $N$ vertices, and the
product runs over the edges of the graph.

Using \eqref{QJsym} the symmetric diagonal form
factors of the current operators are found to be
\begin{equation}
  \label{FJs}
  \FJa(\lan)=
  \left[\sum_{j=1}^N e'(\lambda_j)\right]
    \left[\sum_{j=1}^N q_\alpha(\lambda_j)\right]
  \sum_{\mathcal{T}}  \prod_{l_{jk}\in \mathcal{T}} \varphi_{jk}.
\end{equation}
Supplying these expressions to the expansion theorem \eqref{fftcsa2} one proves the main result
\eqref{main}, as shown in \cite{sajat-currents}.

This proof can be regarded as the finite volume counterpart of the proofs of
\cite{doyon-ghd,dinh-long-takato-ghd}. The works  \cite{doyon-ghd,dinh-long-takato-ghd} operated
directly in the infinite 
volume limit, in the framework of integrable QFT, and results for the currents were obtained through
the so-called LeClair-Mussardo (LM) series. As shown earlier in \cite{sajat-LM}, the LM series
should be regarded as the thermodynamic limit of the expansion theorem \eqref{fftcsa2}.
Thus the new
addition of the work \cite{sajat-currents} was simply to sum up the series \eqref{fftcsa2} in finite
volume, and furthermore to apply and prove it in the Heisenberg chains. 

We remark that essentially the same techniques can also be used in the so-called nested spin chains,
where the fundamental excitations have an inner degree of freedom. In the models related to the
group symmetry $SU(3)$ the finite volume form factors were investigated in
\cite{sajat-artur-nested}, and a formula analogous to  \eqref{fftcsa2} was also derived. It was
briefly explained in \cite{sajat-artur-nested} that the formalism could also be used to derive and
prove results for the currents in nested spin chains. However, the complexity of the method grows
very quickly, and other approaches are more effective in the nested cases.

\section{Algebraic construction}

\label{sec:algebraic}

In this Section we present the algebraic construction of the charge
and current operators in integrable spin chains. The charges were
known since a long time: they can be computed using the so-called
Quantum Inverse Scattering Method (QISM)
\cite{faddeev-how-aba-works,faddeev-history,Korepin-Book}.  However,
 the current operators were only constructed very
recently in \cite{sajat-algebraic-currents}.

The main idea of the QISM is to construct a commuting set of transfer matrices,
which depend on an auxiliary rapidity parameter, and to use their
Taylor expansion around special points to obtain the
charges. Let us therefore consider a generic integrable spin chain with local spaces
$\complex^d$, the total Hilbert space is thus $
\mathcal{H}=\otimes_{j=1}^L \complex^d$.  Let us furthermore take an auxiliary space $V_a$, which we
choose for simplicity to be identical to the physical spaces. Then 
The monodromy matrix acting on $V_a\otimes \mathcal{H}$ is defined as
\begin{equation}
  \label{ttdef}
   T_a(\mu)=\mathcal{L}_{a,L}(\mu)\dots\mathcal{L}_{a,1}(\mu).
\end{equation}
Here $\mathcal{L}_{a,j}(\mu)$, $j=1,\dots,L$ are the rapidity dependent Lax operators
acting on the pair of spaces with indices $a$ and $j$, where $a$
stands for the auxiliary space. 

The transfer matrix is obtained as the partial trace over the auxiliary space:
\begin{equation}
    t(\mu)=\text{Tr}_a  T_a(\mu).
\end{equation}

We require that the following exchange relation holds for the local
Lax operators:
\begin{equation}
  \label{RLL}
  \begin{split}
  R_{b,a}(\nu,\mu) & \mathcal{L}_{b,j}(\nu)  \mathcal{L}_{a,j}(\mu)= \mathcal{L}_{a,j}(\mu) \mathcal{L}_{b,j}(\nu)    R_{b,a}(\nu,\mu)
\end{split}
\end{equation}
with $a,b$ referring to two different auxiliary spaces.  Here
$R_{b,a}(\nu,\mu)$ is the so-called $R$-matrix acting on two auxiliary
spaces. Consistency requires that the $R$-matrix satisfies the
so-called 
Yang-Baxter relation:
\begin{equation}
  \label{YB}
  \begin{split}
     R_{12}(\lambda_{1},\lambda_2)&R_{13}(\lambda_1,\lambda_3)R_{23}(\lambda_2,\lambda_3)=\\
&  =R_{23}(\lambda_2,\lambda_3) R_{13}(\lambda_1,\lambda_3) R_{12}(\lambda_{1},\lambda_2).
  \end{split}
\end{equation}
This is a relation for operators acting on the triple tensor product $V_1\otimes V_2\otimes V_3$
and we assume $V_j\simeq \complex^d$. It is understood that each $R_{jk}$ acts only on the
corresponding vector spaces. The Yang-Baxter equation can be
understood as a consistency equation for factorized elastic
scattering, if the $R$-matrix is interpreted as a scattering matrix \cite{Mussardo-review}.
Examples for $R$-matrices  can be found in
\cite{Baxter-Book,Korepin-Book,Hubbard-Book}. In many cases
$R(\mu,\nu)$ is of difference form, i.e. it only depends on the
combination $\mu-\nu$. However, this is not the only possibility, and
we allow for a generic dependence on both rapidity parameters.

It follows from \eqref{YB} that  $\mathcal{L}_{a,j}(\mu)=R_{a,j}(\mu,\xi_0)$
is a solution to \eqref{RLL}, where $\xi_0$ is some parameter of the model. In the following we
use this choice and we assume simply $\xi_0=0$.

The fundamental exchange relations \eqref{RLL} imply that
\begin{equation}
[t(\mu), t(\nu)]=0.
\end{equation}
The Taylor expansion of the transfer matrices are used to construct
the canonical charges. Let us assume that the so-called regularity and unitarity conditions hold:
\begin{equation}
  \label{inv}
  \begin{split}
    R(\lambda,\lambda)&=P\\
  R_{12}(\lambda_1,\lambda_2)R_{21}(\lambda_2,\lambda_1)&=1.
  \end{split}
 \end{equation}
Here  $P$ is the permutation operator and $R_{21}(u,v)=PR_{12}(u,v)P$.

Then a generating function for global charges is defined as \cite{Korepin-Book,faddeev-history}
\begin{equation}
  \label{Qnudef}
  Q(\nu)\equiv (-i) t^{-1}(\nu)\frac{d}{d\nu} t(\nu).
\end{equation}
This definition agrees with the generating function defined earlier in
\eqref{qnudef}, thus the canonical local charges $Q_\alpha$ are given by its expansion coefficients. 
The $Q_\alpha$ are extensive, and the density
$q_\alpha(x)$ spans $\alpha$ sites \cite{Luscher-conserved}; for example $H\sim Q_2$. The definition
\eqref{Qnudef} makes sense in any finite volume, but it gives 
the correct $Q_\alpha$ if the charge fits into the volume,
i.e. $L>\alpha$. In the $L\to\infty$ limit the operator $Q(\nu)$
is expected to be quasi-local, at least in some neighborhood of $\nu=0$. For proofs in concrete cases see
\cite{prosen-xxx-quasi,prosen-enej-quasi-local-review,sajat-su3-gge}. 

In practical computations the inverse of the transfer matrix can be replaced by the space reflected
TM. Defining
\begin{equation}
  \label{tbardef}
\bar  T_a(\mu)=\mathcal{L}_{a,1}(\mu)\dots\mathcal{L}_{a,L}(\mu),\qquad
  \bar  t(\mu)=\text{Tr}_a \bar  T_a(\mu)
\end{equation}
it can be shown that the following asymptotic inversion holds:
\begin{equation}
  t(\lambda)\bar t(-\lambda)=1+\ordo(e^{-cL}),
\end{equation}
This relation holds at least in some neighborhood of $\lambda=0$, such that the decay constant $c>0$ depends on
$\lambda$. For 
concrete proofs see \cite{prosen-xxx-quasi,prosen-enej-quasi-local-review,sajat-su3-gge}. 

Let us now turn to the current operators. It was shown in
\cite{sajat-algebraic-currents} that they can be embedded into the
QISM framework, which we now explain.
 
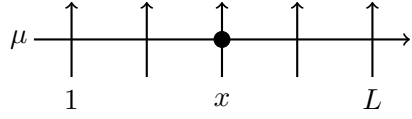
\begin{figure}[t]
  \centering
  \begin{tikzpicture}
    \draw[thick,->] (0,0) -- (5,0);
    \foreach \x in {0,1,2,3,4}    \draw[thick,->] (\x + 0.5,-0.5) -- (\x + 0.5,0.5);
    \draw[fill] (2.5,0) circle (3pt);
     \node at (-0.2,0) {$\mu$};
        \node at (0.5,-0.8) {$1$};
     \node at (2.5,-0.8) {$x$};
        \node at (4.5,-0.8) {$L$};
   \end{tikzpicture}
  \caption{The generating function $q(\mu,x)$ for the charge densities is
    obtained by taking a derivative with respect to the rapidity at a
    single site. Here periodic boundary conditions are understood. The
  crossing with the dot denotes the action of $-i \partial_\mu R(\mu,0)$.}
  \label{fig:TLu}
\end{figure}

First we find the operator density for the generating function $Q(\mu)$. 
Writing $Q(\mu)=\textstyle \sum_{x=1}^L   q(\mu,x)$ we can identify
\begin{equation}
  \label{qmuxdef}
  \begin{split}
  q(\mu,x)&\equiv (-i) t^{-1}(\mu) \text{Tr}_a    \left[  T_a^{[L,x+1]}(\mu)    
      \partial_\mu {\mathcal{L}}_{a,x}(\mu)    T_a^{[x-1,1]}(\mu) \right].
  \end{split}
\end{equation}
Here we defined the partial monodromy matrices acting on a segment $[x_1\dots x_2]$
as
\begin{equation}
  T^{[x_2,x_1]}_a(\mu)=\mathcal{L}_{a,x_2}(\mu) \dots \mathcal{L}_{a,x_1}(\mu).
\end{equation}
A graphical representation of the generating function is given in Fig. \ref{fig:TLu}.

We also define a generating function for the current operators. Naturally this will be a function of
two auxiliary variables and a space coordinate. It is defined
implicitly through the continuity relation \eqref{Jmunudef}.
It is our goal to derive explicit formulas for $J(\mu,\nu,x)$.

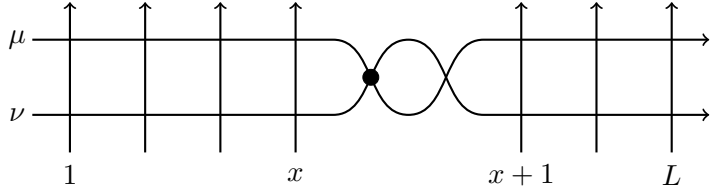
\begin{figure}[t]
  \centering
  \begin{tikzpicture}
          \node at (-0.2,1) {$\mu$};
               \node at (-0.2,0) {$\nu$};
    \draw[thick,->] (0,0) to (4,0) to[out=0,in=180] (5,1)
    to[out=0,in=180] (6,0) to (9,0);
    \draw[thick,->] (0,1) to (4,1)  to[out=0,in=180] (5,0)
    to[out=0,in=180] (6,1) to (9,1);
      \draw[fill] (4.5,0.5) circle (3pt);
\foreach  \x in {0,1,2,3,6,7,8} \draw[thick,->] (\x + 0.5,-0.5) to (\x + 0.5,1.5);
     \node at (0.5,-0.8) {$1$};
     \node at (3.5,-0.8) {$x$};
        \node at (6.5,-0.8) {$x+1$};
        \node at (8.5,-0.8) {$L$};
\end{tikzpicture}
  \caption{The operator $\Omega(\mu,\nu,x)$ positioned at site $x$. Here periodic boundary
conditions are assumed in the horizontal direction. As before, the dot
denotes derivative with respect to a rapidity parameter. Altogether the
operator insertion in the middle is given by eq. \eqref{opins}.}
  \label{fig:Om}
\end{figure}

First it can be shown using a repeated action of the Yang-Baxter equation that the solution of the
analogous operator equation 
\begin{equation}
  \label{opop3c}
 t^{-1}(\nu)  \left[t(\nu), q(\mu,x)\right]=
  \Omega(\mu,\nu,x)- \Omega(\mu,\nu,x-1)
\end{equation}
is given by 
\begin{equation}
  \label{Omegadef}
  \begin{split}
   \Omega(\mu,\nu,&x)=
   t^{-1}(\nu) t^{-1}(\mu) \text{Tr}_{ab}\left[  T_a^{[L,x+1]}(\mu)
     T_b^{[L,x+1]}(\nu) \mRR_{a,b}(\mu,\nu)    T_a^{[x,1]}(\mu)   T_b^{[x,1]}(\nu)\right].
  \end{split}
\end{equation}
 where $a$ and $b$ are two different auxiliary spaces and
 \begin{equation}
   \label{opins}
  \mRR_{a,b}(\mu,\nu)= (-i)  R_{b,a}(\nu,\mu) \partial_\mu   R_{a,b}(\mu,\nu).
\end{equation}
is an operator insertion acting only on the auxiliary spaces.
A pictorial representation of $\Omega(\mu,\nu,x)$ can be found
in Fig. \ref{fig:Om}. Note that apart from the $t^{-1}$ operators $\Omega$ is a ``double row'' matrix product
operator, and the only difference as opposed to a simple product of two transfer matrices is the insertion
$\mRR_{a,b}(\mu,\nu)$, which couples two monodromy matrices. 

Taking a $\nu$-derivative on the l.h.s. of \eqref{opop3c} we recognize the continuity equation
\eqref{Jmunudef} and identify
\begin{equation}
  \label{JmunuabOmega}
  J(\mu,\nu,x)=- t(\nu) \partial_\nu \Omega(\mu,\nu,x-1)  t^{-1}(\nu).
\end{equation}
Let $\ket{\Psi}$ be an arbitrary eigenstate of the commuting transfer matrices.
For the mean values we get:
\begin{equation}
  \label{Jomega}
   \bra{\Psi} J(\mu,\nu,x)\ket{\Psi}=-\partial_\nu    \bra{\Psi} \Omega(\mu,\nu,x-1)\ket{\Psi}.
 \end{equation}
This connects the $\nu$-derivatives of $\Omega(\mu,\nu,x)$ to the current mean values. To complete
the picture, it can be shown that the initial value at $\nu=0$ is given by
$\Omega(\mu,0,x)= q(\mu,x)$.
Thus $ \Omega$ not only describes all (generalized) currents, but also all charge
densities.

It was shown in \cite{sajat-algebraic-currents} that $\Omega(\mu,\nu)$ is symmetric with
respect to its variables iff the $R$-matrix is of the difference form:
\begin{equation}
  R(\mu,\nu)=R(\mu-\nu).
\end{equation}
Examples for this are the fundamental $SU(N)$-invariant models and the XXZ and XYZ chains. A famous
counter-example is the one dimensional Hubbard model.

The mean values of $\Omega(\mu,\nu)$ can be obtained with a trick that was originally developed in
\cite{XXZ-finite-T-factorization}: it can be shown that the mean values are related to a
transfer matrix eigenvalue in an 
auxiliary problem, namely in an enlarged spin chain with two extra sites.  This is a rather
technical computation, and we do not discuss its details.

Let us however present the end result in the Heisenberg spin chains:
it was shown in \cite{sajat-algebraic-currents} that 
\begin{equation}
  \label{omegapsi}
  \bra{\lan}\Omega(\mu,\nu)\ket{\lan}=\Psi(\mu,\nu)+\dots,
\end{equation}
where $\Psi(\mu,\nu)$ is 
\begin{equation}
  \label{psimunu}
  \begin{split}
\Psi(\mu,\nu)= {\bf h}(\nu) \cdot G^{-1}\cdot {\bf h}(\mu),
  \end{split}
\end{equation}
where ${\bf h}(\mu)$ is a vector with components given by $h_2(\lambda_j-\mu)$ with the $h_2$ function defined in
\eqref{halpha0} (similarly for  ${\bf h}(\nu)$),
 and the dots signal correction terms that decay for small $\mu,\nu$ at least as $\mu^L$ or $\nu^L$.

Looking at the connection \eqref{Jomega}, and considering the expansion of $\Psi(\mu,\nu)$ we can see
that the statement \eqref{main} is indeed reproduced by this computation. The correction terms
mentioned above do not influence the final results for those generalized currents $J_{\alpha,\beta}$, which fit into a
given volume, because their mean values are obtained by taking derivatives of $\Psi(\mu,\nu)$ an appropriate times at 
$\mu,\nu=0$.

\section{Perturbation theory and long range deformations}

\label{sec:longrange}

A surprising new interpretation of the current operators was given in
\cite{sajat-longcurrents}: It was shown that they trigger a
specific type of integrable long range deformation of spin
chains, that were studied earlier in the context of the AdS/CFT correspondence
\cite{beisert-dilatation,beisert-long-range-1,beisert-long-range-2}. Let us now summarize this connection.

We consider first the infinite volume situation, and perform a continuous deformation on our model, such that
$\kappa$ is a deformation parameter. We require that the deformation
preserves integrability, thus we look for a $\kappa$-dependent set of
operators $\{Q_\alpha^\kappa\}_{\alpha=1,2,\dots}$, such 
that
\begin{equation}
  \label{kappacomm}
  [Q_\alpha^\kappa,Q_\beta^\kappa]=0,
\end{equation}
The initial condition should be given by
$Q_\alpha^{\kappa=0}=Q_\alpha$ with $Q_\alpha$
being the original charges. We require that the resulting operators
are extensive with an operator density defined as
\begin{equation}
  \label{Qxdef2}
  Q_\alpha^\kappa=\sum_{x=-\infty}^\infty q^\kappa_\alpha(x).
\end{equation}
It is useful to think about the deformation as a formal power series in $\kappa$, which is assumed to be
convergent for the operator density at least in some neighborhood of $\kappa=0$:
\begin{equation}
  \label{defkifejt}
  Q_\alpha^\kappa=\sum_{n=0}^\infty \frac{\kappa^n}{n!}Q_\alpha^{(n)}.
\end{equation}
Then the commutativity condition \eqref{kappacomm} can be analyzed order by order in $\kappa$.

The locality properties of the charges typically change under the deformation: the charge densities
acquire contributions with increasing range. We can expect that the resulting charges are still
quasi-local (for definitions and properties see \cite{prosen-enej-quasi-local-review}).

One possibility to construct such long range deformations is by postulating a
generating equation for the charges, that describes ``evolution in
$\kappa$''. This takes the Lax form \cite{beisert-long-range-2}
\begin{equation}
  \label{defdef}
\frac{d}{d\kappa} Q_\alpha^\kappa= i[X(\kappa),Q_\alpha^\kappa].
\end{equation}
Here $X(\kappa)$ is a formal operator, which is chosen in a specific way such that the solution of
\eqref{defdef} is a quasi-local operator. The generating equation can be understood as an iterative
equation once we expand it into a Taylor series in $\kappa$.

Any such deformation leaves the commutation relations between the charges
unmodified, which follows from the formal Jacobi identity:
\begin{equation}
  \frac{d}{d\kappa} [Q_\alpha^\kappa,Q_\beta^\kappa]=
  i[X(\kappa), [Q_\alpha^\kappa,Q_\beta^\kappa]].
\end{equation}
Thus commutativity is preserved.

It was explained in \cite{beisert-long-range-2} that
there are essentially two different types of generating operators yielding non-trivial deformations,
the so-called boost type and bi-local type generators. Let us focus on the boost type of
deformations. In this case
  \begin{equation}
    \label{Xboost}
    X(\kappa)=-  \sum_x x q_\alpha^\kappa(x).
  \end{equation}
It was a surprising result of \cite{sajat-longcurrents}, that in this case the deforming operators
are actually the current operators. To see this, let 
 us also define the deformed current densities as
  \begin{equation}
  \label{Jabkappa}
i  \left[Q_\beta^\kappa,  q_\alpha^\kappa(x)\right]=J^\kappa_{\alpha,\beta}(x)-J^\kappa_{\alpha,\beta}(x+1).
\end{equation}    
A simple rewriting of the deformation equation \eqref{defdef} using \eqref{Xboost}-\eqref{Jabkappa} gives
  \begin{equation}
    \label{qabjabk}
    \dperdk Q_\beta^\kappa=\sum_{x}  J_{\alpha,\beta}^\kappa(x),
  \end{equation}

This relation leads to the possibility of finding the current mean values using the deformation. To
see this, let us assume that we can define the deformation also in finite volume, at least up to
some fixed order in $\kappa$. Assuming further that we can also compute the eigenvalues  of the
charges as a function of $\kappa$, the current mean values of the original model will be given by
the first order correction:
\begin{equation}
  \label{fgh}
  \vevv{J_{\alpha,\beta}(x)}=\frac{1}{L}\left.\frac{d \Lambda_\beta^\kappa}{d\kappa}\right|_{\kappa=0},
\end{equation}
where $\Lambda_\beta^\kappa$ is the $\kappa$-dependent eigenvalue of the deformed charge $Q_\beta^\kappa$.

The problem with this approach is that it is not known how to perform
the long range deformation in finite
volume.
The difficulty lies in the ``wrapping problem'': the deformed
operators acquire contributions with arbitrary length for any fixed
$\kappa$, and generally it is not known how to ``wrap'' these
operators around the finite volume.
Existing approaches claim that the deformation can be performed at least up to some order in
$\kappa$, as long as the relevant lower order contributions to the
operators still fit into the given volume. In this case reliable results are found using the ``Asymptotic
Bethe Ansatz''. The idea is to compute the deformation of eigenstates
in infinite volume, to extract the 
propagation and scattering phases, and to use this information even in finite volume, to set up 
approximate Bethe Ansatz equations. We now summarize this approach.

First we return to the infinite volume and consider an eigenstate
$\ket{\Psi^0}$ of the un-deformed theory.
Then we define the deformation of the state as
\begin{equation}
  \label{psidefdef}
  \frac{d}{d\kappa} \ket{\Psi^\kappa}=-iX(\kappa) \ket{\Psi^\kappa}
\end{equation}
with the initial condition $\ket{\Psi(\kappa=0)}=\ket{\Psi^0}$. The
eigenstates are mapped to eigenstates and the charge eigenvalues
$\Lambda_\alpha^0$ are not deformed: 
\begin{equation}
  Q_\alpha^\kappa\ket{\Psi^\kappa}=\Lambda_\alpha^0 \ket{\Psi^\kappa}.
\end{equation}
However, it follows from the particular form \eqref{Xboost} that the lattice momentum is changed according to
\begin{equation}
 \dperdk P(\kappa)=\Lambda_\alpha.
\end{equation}
As it was discussed above, the eigenvalues of the charges do not
change under the deformation, thus we get
the simple solution
\begin{equation}
  \label{Pelsodef}
  P(\kappa)=P+\Lambda_\alpha \kappa.
\end{equation}
This is the generalization of the standard boost operation known in
Lorentzian of Galilean invariant models. It holds for all eigenstates
of the infinite volume system.

On one-particle states the deformation results in the modified
momentum-rapidity relation
\begin{equation}
  p(\lambda)\quad\to\quad p(\lambda)+\kappa h_\alpha(\lambda).
\end{equation}
The quasi-locality of the deformation implies that the
multi-particle momentum of Bethe states has to be of the form
\begin{equation}
  \label{pdefdef}
  P^\kappa=\sum_{j=1}^N p^\kappa(\lambda_j),\qquad
  p^\kappa(\lambda_j)=p(\lambda)+\kappa h_\alpha(\lambda_j).
\end{equation}
Furthermore it can be argued that the scattering phase in the Bethe Ansatz is not changed when
viewed as a function of the rapidities. 

Let us now put the theory into finite volume. The assumption of the asymptotic Bethe Ansatz is that
up to a given order in $\kappa$ we can use the infinite volume data to write down the Bethe states
and the corresponding Bethe equations. These will determine the finite volume rapidities. We get
\begin{equation}
  \label{bethedeformed}
  p^\kappa(\lambda_j)L+\sum_{k\ne j} \delta(\lambda_j-\lambda_k)=2\pi I_j,
\end{equation}
where $I_j$ are integer quantum numbers, which have to be kept fixed
by continuity. Then the $\kappa$-dependence of the charge eigenvalues
comes simply from the $\kappa$-dependence of the solution to this set of equations. 

The first order correction to the rapidities is easily computed by taking derivatives of the above equations,
leading to
\begin{equation}
  \frac{\partial \lambda_j}{\partial \kappa}=LG^{-1} {\bf h}_\alpha,
\end{equation}
where ${\bf h}_\alpha$ is a vector with elements
$h_\alpha(\lambda_j)$. Computing finally \eqref{fgh} gives
\begin{equation}
  \label{poi}
    \vevv{J_{\alpha,\beta}(x)}=\sum_{j=1}^N h'_\beta(\lambda_j) \frac{\partial \lambda_j}{\partial \kappa}=
{\bf h'_\beta} G^{-1} {\bf h}_\alpha.
\end{equation}
Thus we have obtained one more derivation of the formula \eqref{main}.

The reader might wonder to what extent this argument can be considered rigorous. It was argued in
\cite{sajat-longcurrents} that the calculation can be trusted as long as the corresponding operators
fit into a particular volume.
However, there is one specific case where the computation can be made
completely rigorous: the case when the
deforming charge is chosen simply as the conserved spin operator. To be precise we choose
 \begin{equation}
    \label{Xboostspin}
    X(\kappa)=-  \sum_x x S^z(x).
  \end{equation}
This deformation generates a homogeneous twist along the chain, nevertheless preserving complete
integrability to all orders in $\kappa$ \cite{Dzyaloshinsky-Moriya-xxz}. The deformed Bethe
equations \eqref{bethedeformed} hold 
exactly with
\begin{equation}
    p^\kappa(\lambda)=p(\lambda)-\kappa.
\end{equation}
The $\kappa$-derivative of the deformed Hamiltonian is the spin-current, and one obtains the
corresponding mean value formula \footnote{We acknowledge collaboration on this computation with
  Lorenzo Piroli.}. This twisting procedure underlies the proof of
\cite{kluemper-spin-current} for the spin current, which was given
prior to the finite volume results of the present authors.

\subsection{Weak thermalization}

There is an interesting implication of the computations above: The
current operators lead to a weak breaking of integrability.

Let us fix the index $\beta$, and truncate the power series 
\eqref{defkifejt} to the first order. Then we find that the operators
\begin{equation}
  Q_\alpha+\kappa\sum_x J_{\beta,\alpha}(x),\qquad \alpha=2,3,\dots
\end{equation}
commute up to order $\kappa$. For the case of $\alpha=2$ we get the deformation of the Hamiltonian,
which reads
\begin{equation}
  H+\kappa \sum_x J_\beta(x),
\end{equation}
with $J_\beta(x)$ being the physical current of the charge $Q_\beta$. The computations above imply
that this perturbation does not break integrability at the leading order.

The same statement was also found in two different settings using different methods.

The work
\cite{doyon-weak-breaking} considered the problem of weak thermalization in cases when an integrable
Hamiltonian is perturbed by some local operator. It was found that for small perturbations of order
$\kappa$ there is thermalization
on the time scale of $\kappa^{-2}$, if the perturbing operator is generic.  However, for a special
class of operators thermalization only happens on longer time scales. Interestingly, it was found in
\cite{doyon-weak-breaking} that the current operators belong to this
special class. However, this class of local operators is characterized by completely different tools, 
and at present it is not known how to connect the formalism of \cite{doyon-weak-breaking}  to that of
\cite{sajat-longcurrents}. 

Finally, the recent work \cite{sajat-levelspacing} studied the level spacing distribution in perturbed integrable
models. It is known that integrable models show Poisson distribution, whereas in the non-integrable cases we expect
Wigner-Dyson distribution due to level repulsion. In the presence of integrability breaking one expects a crossover
between the two distributions: from Poisson in intermediate volumes to Wigner-Dyson in the thermodynamic limit. Such a
crossover was observed for generic integrability breaking in \cite{levelspacing-crossover}. The work
\cite{sajat-levelspacing} found that if 
the perturbation is triggered by one of current operators, then larger volumes and/or larger couplings are needed for
the crossover. In particular, the critical coupling needed to cross over to the Wigner-Dyson statistics scales
differently with the volume, signaling 
once again that the current operators do not break integrability at the leading order.

\section{Implications for correlation functions}

\label{sec:corr}

\subsection{Factorized correlation functions}

\label{sec:factorized}

In the papers \cite{sajat-currents,sajat-algebraic-currents} an
interesting connection was pointed out to the theory of factorized
correlation functions.
This theory was developed in a series of works with contributions
from many researchers
\cite{hgs1,hgs2,HGSIII,bjmst-fact4,bjmst-fact5,bjmst-fact6,boos-goehmann-kluemper-suzuki-fact7,XXXfactorization}.
The history of the theory started with the work \cite{boos-korepin-first-factorization} of Boos and Korepin
who observed that some concrete multiple integral formulas for correlations of the XXX chain can be
factorized (see also \cite{boos-korepin-smirnov-fact2}). Afterwards this was developed into a full
algebraic theory, which deals with the normalized mean values of the form
\begin{equation}
  \bra{\Psi}\ordo\ket{\Psi},
\end{equation}
where $\ordo$ is a short range operator of the Heisenberg chain (the XXZ and XXX cases need to be
treated separately), and $\ket{\Psi}$ is an eigenvector of the commuting set of transfer
matrices. The works cited above only treated the ground state in
finite and infinite volume, and also finite temperature cases, and
the extension to arbitrary excited states was performed in \cite{sajat-corr,sajat-corr2}.

The main statement of the theory is that each mean value can be expressed using just a few functions.
To be more precise, the mean values are expressed as combinations of the
Taylor coefficients of some functions with one or two auxiliary variables. The number of functions depend on
the model: In the XXZ chain it is enough to consider just two functions (commonly denoted as
$\omega(\mu,\nu)$ and $\omega'(\mu,\nu)$) for operators that are symmetric with respect to spin-flip.
For generic operators an additional function $\varphi(\mu)$ is also needed. In the XXX case the
situation is more involved: for operators that are $SU(2)$-symmetric (or for states with zero
magnetization) it is enough to consider just a
single function $\omega(\mu,\nu)$, whereas the full solution in the generic case is not yet
known. We note that even the XYZ model was treated in \cite{xyz-factorization} and it is conjectured
that three functions with two variables are enough to describe all correlation functions in that model.

We now give a flavor for the results for the XXX chain. Let us consider
states with zero total magnetization. Then short range $z-z$ correlations can be expressed as \cite{XXXfactorization}
\begin{align}
\label{sig13}
    \vev{\sigma_{1}^z\sigma_2^z}=&
\frac{1}{3}(1-\Psi_{0,0})\\
   \vev{\sigma_{1}^z\sigma_3^z}
=&\frac{1}{3}(1-4\Psi_{0,0}+\Psi_{1,1}-\frac{1}{2}\Psi_{2,0})\\
\label{sig14}
\begin{split}
  \vev{\sigma_{1}^z\sigma_4^z}=&
\frac{1}{108} (288 \Psi_{1, 1}   - 15 \Psi_{2, 2} +   10
\Psi_{3, 1}+
2 \Psi_{0, 0} (-162  - 42 \Psi_{1, 1}  +   3 \Psi_{2, 2}- \\
&- 2 \Psi_{3, 1})+ \Psi_{2, 0}( - 156 + 12 \Psi_{1, 1}  - 6 \Psi_{2, 0})+
\Psi_{1,0}(84 \Psi_{1, 0} - 12 \Psi_{2, 1}+ \\ &+   4 \Psi_{3, 0})+36  ),
\end{split}
\end{align}
where $\Psi_{\alpha,\gamma}$ \footnote{In this Section we do not use $\beta$ as an index in order to reserve it for the inverse temperature} are Taylor coefficients of a function
$\Psi(\mu,\nu)$, which is related to the above mentioned
$\omega(\mu,\nu)$ in a linear way. The expansion reads
\begin{equation}
 \Psi_{\alpha,\gamma}=\partial_{\mu}^\alpha \partial_{\nu}^\gamma \Psi(\mu, \nu)|_{\mu,\nu=0}.
\end{equation}
 For our purposes the function $\Psi(\mu,\nu)$ is more
convenient. 

In \cite{sajat-corr2} it was argued that the function $\Psi(\mu,\nu)$
is given for excited states by \eqref{psimunu}. However, we also explained
that this is equal to the mean value of the currents.
Thus, we find that  the generalized current operators are those special operators whose mean values are
  linear in the co-efficients of $\Psi(\mu,\nu)$. This was not clear
  in the series of works \cite{hgs1,hgs2,HGSIII,bjmst-fact4,bjmst-fact5,bjmst-fact6,boos-goehmann-kluemper-suzuki-fact7,XXXfactorization}.

Furthermore, it is possible to give a rather direct connection between the operator
$\Omega(\mu,\nu)$ introduced in Section \ref{sec:algebraic} and the theory of
factorized correlation functions. It was shown in \cite{sajat-algebraic-currents} that if we define
$\Omega(\mu,\nu)$ for an {\it inhomogeneous chain} with a formula analogous to \eqref{Omegadef}, and then specify the 
parameters $\mu,\nu$ to two selected inhomogeneities, then actually we obtain a certain component of
the corresponding two-site reduced density matrix. To be more precise, the following can be proven
in a straightforward way:
\begin{equation}
  \label{xi1xi2}
  \Omega(\xi_1,\xi_2,x=2)=\mRR_{1,2}(\xi_1,\xi_2),
\end{equation}
where now the operator $\mRR_{1,2}(\xi_1,\xi_2)$ acts on the physical spaces 1 and 2.
This means that for these special values $\Omega(\mu,\nu,x)$ becomes an ultra-local operator acting only
on the first two sites. 
This is the direct  bridge to the theory of factorized correlation functions.
The result \eqref{xi1xi2} is analogous to the 
``solution of the inverse problem'' 
\cite{goehmann-korepin-inverse,maillet-terras-inverse}, where the monodromy matrix elements can be
specialized such that they become ultra-local operators acting on single sites only.

\subsection{Asymptotic behaviour of dynamical correlations}

Here we present a new application of the results for correlation
functions to GHD. We consider the asymptotic behaviour of correlation
functions in a dynamical situation. These objects were considered in
\cite{Doyon-GHD-LM}, where a rather general formula was derived based
on the theory of hydrodynamic projections (for an alternative treatment see \cite{axel-milosz-javulo}).  The most
generic formula for 
the two-point functions concerns spatially inhomogeneous but slowly
varying initial configurations, and their correlation functions in the
long time and large distance limits.

Here we focus on a more
restrictive setup, and we consider dynamical correlation functions in
spatially homogeneous equilibrium states. The most general result of
\cite{Doyon-GHD-LM} can be restricted to such cases, which means that
we get a hydrodynamic prediction for the dynamical correlations in
equilibrium. This prediction involves certain dynamical amplitudes
associated to the operators, and below we show that these amplitudes
can be computed from the current mean values, using the general theory
of factorized correlation functions. It is important that we do not
prove the formulas of \cite{Doyon-GHD-LM}, we just merely apply them
and compute the amplitudes that appear in them. We will see that the
amplitudes also show a specific type of factorization. The results
within this Subsection are new. We should note that the dynamical
correlation functions of the spin chain are subject of active research
\cite{xx-uj-1,xx-uj-2,karol-frank-junji-xx-uj-3,karol-frank-junji-xxz-uj-1,karol-frank-junji-xxz-uj-2}, thus it is likely that the predictions presented below will be
proven or disproven in specific situations. 

Let us then consider an equilibrium state $\ket{\Omega}$ in the XXX
Heisenberg spin chain, which is described by the Bethe root
densities. We intend to compute the asymptotic limit of correlation functions
\begin{equation}
  \bra{\Omega}O_1(x,t)O_2(0,0)\ket{\Omega}.
   \label{eq:correlation_def}
\end{equation}
We consider the asymptotic limit of this correlation function along a ray, i.e. we take  $x,t\rightarrow\infty$ with
$x/t=\xi$ fixed. In this limit the correlation function (\ref{eq:correlation_def}) becomes \cite{Doyon-GHD-LM}: 
\begin{equation}
  \bra{\Omega}O_1(x,t)O_2(0,0)\ket{\Omega}=t^{-1}
  \sum_{j=1}^\infty\sum_{\lambda\in\Lambda_j(\xi)}\frac{\rho_j(\lambda)w_j(\lambda)}{|(v_j^{\textrm{eff}})'(\lambda)|}V_j^{O_1}(\lambda)V_j^{O_2}(\lambda).
\end{equation}
Here the summation over $j$ runs over the particle species,  $\rho_j(\lambda)$ are the root densities,
$v_j^{\textrm{eff}}(\lambda)$ are the effective velocities and $\Lambda_j(\xi)$ is the set of rapidities for which
$v_j^{\textrm{eff}}(\lambda)=\xi$. Furthermore, $w_j(\lambda)$ are one more set of thermodynamic functions described
below. Finally, $V_j^{O_1}(\lambda)$ and $V_j^{O_2}(\lambda)$ are certain amplitudes associated to the local operators,
which can be interpreted as thermodynamical form factors.

These amplitudes can be computed using the theory of hydrodynamic projections
\cite{Doyon-GHD-LM,doyon-hydro-projection}. For the discussion of this theory we refer the reader to
\cite{Doyon-GHD-LM,doyon-hydro-projection}; here we just apply the formalism and show how the current mean values enter
the operator amplitudes $V_j^{O_1}(\lambda)$ and $V_j^{O_2}(\lambda)$ for generic operators $O_1$ and $O_2$.

As discussed in \cite{Doyon-GHD-LM}, the operator amplitudes are computed by taking derivatives of the mean values of
the operators with respect to
parameters of the equilibrium ensemble. These parameters can be chosen for example as Lagrange multipliers of the
GGE.

We explained above that the mean values of the XXX chain are factorized in the sense that they can be expressed as a sum
of products of simple building blocks $\Psi_{\alpha,\gamma}$. We write this statement formally as
\begin{equation}
    \expval{\mathcal{O}} = \xi_0 + \sum_{j=1}^n \xi_j \prod_{k=1}^{m_j} \Psi_{\alpha_{j,k},\gamma_{j,k}} \hspace{1.5 cm} \xi_j\in \mathbb{R},
  \end{equation}
  where $\xi_j$ are numerical coefficients, such as those present in \eqref{sig13}-\eqref{sig14}.

The derivatives with respect to GGE parameters act on such a product by the Leibniz rule. If we define the amplitudes $V_j^{\Psi_{\alpha,\gamma}}$ for the building blocks then for any operator we have
\begin{equation}
    V_j^{\mathcal{O}}(\lambda) = \sum_{j=1}^n \xi_j \sum_{k=1}^{m_j} V_j^{\Psi_{\alpha_{j,k},\gamma_{j,k}}}(\lambda) \prod_{\substack{l=1\\l \neq k}}^{m_j} \Psi_{\alpha_{j,k},\gamma_{j,k}},
\end{equation}
which is merely the application of the Leibniz rule. The remaining task  is to determine the functions
$V_j^{\Psi_{\alpha,\gamma}}$. In Appendix \ref{sec:proj} we show that they are given for every $\alpha,\gamma$ by
  \begin{equation}
    V_j^{\Psi_{\alpha,\gamma}}(\lambda) = \frac{1}{\pi} \frac{\big( h_{\alpha+1,j}' \big)^\text{dr}(\lambda) h_{\gamma+2,j}^\text{dr}(\lambda)}{\rho_j^t(\lambda)}.
    \label{eq:building_block_Vfactor}
\end{equation}
This result is computed from our new formulas for the current mean values. Together with the algebraic part of the
factorization procedure they yield the operator amplitudes for the asymptotic correlation functions.

\section{Conclusions and Outlook}

\label{sec:conclusions}

In this review we treated the current operators of integrable
models. We showed that their mean values take a remarkably simple
form, both in finite volume and in the thermodynamic limit.

This remarkable simplicity is tied to the two-body irreducibility of the Bethe
Ansatz wave function. Despite the simplicity of the final formulas,
the actual proofs require relatively large amount of work. The
situation is 
similar to the problem of the norm of the Bethe Ansatz wave function:
Even though the end result is always simple (it is given by the Gaudin
determinant), the actual proofs are not straightforward.

Let us now summarize the advantages and disadvantages of the three
different proofs that were discussed in this review:
\begin{itemize}
\item {\bf Form factor expansion.} This method is rather general, it
  can be applied both in spin chains and in continuum models, be it
a  relativistic QFT \cite{dinh-long-takato-ghd} or a non-relativistic models such as the
Lieb-Liniger model. A disadvantage is that the graph theoretical
summations need a dedicated analysis, which can become overly
complicated in multi-component models.
\item {\bf Algebraic construction.} This seems to be the most general
  and most straightforward method for spin chains, and it is
  completely rigorous. Its applicability
  is much wider than the cases considered here. It gives a technical
  explanation as to why a simple final result can exist: because the
  mean value in question is tied to a transfer matrix eigenvalue in an
  auxiliary problem. However, the physical meaning of this relation is
  not clear. A disadvantage of the method is that in the present form
  it is limited to lattice models. An extension to continuous models
  is an exciting future direction.
\item {\bf Long range deformations.} This is perhaps the easiest
  method: Once the general recipe is understood, it can be applied to
  a large class of models with a very straightforward computation. A
  drawback is that the method is not rigorous,
  because the statements behind the asymptotic Bethe Ansatz are not
  rigorously established.
\end{itemize}

There are a number of open questions that deserve further study. We
list some of these questions below:

\begin{itemize}
\item {\bf Current mean values in the Hubbard model.} The Hubbard
  model is one of the most important models of interacting 1D
  fermions. Its algebraic formulation is known, it involves the famous
  $R$-matrix of Shastry \cite{Hubbard-Book}, which is not of
  difference form. The current operators in
  this model have not yet been studied, even though there are works
  treating various aspects of GHD in this model
  \cite{jacopo-enej-hubbard,hubbard-ghd-2,hubbard-ghd-3}. It would be
  desirable to prove the corresponding formulas for the current mean values.
\item {\bf Currents in models without particle conservation.} So far
  GHD was limited to integrable models with global $U(1)$-symmetry,
  leading to particle number conservation. However, the flow equations
  for the charges can be formulated even in the absence of particle
  conservation. One example for such a model is the XYZ spin chain,
  which is integrable and which belongs to the class of models treated
  by the QISM framework. Eigenstates in that model can be constructed
  using a generalized Algebraic Bethe Ansatz, which mixes sectors with
  different particle numbers. It would be desirable to work out the
  current mean values in this model, which would lead to a GHD setup
  without fundamental particle conservation.
\item {\bf Long range deformations.} The work \cite{sajat-longcurrents}
  treated the long range deformations to first order in the
  deformation parameter $\kappa$. It is an interesting question,
  whether there are long range vertex models which could accommodate
  even higher order corrections. This could lead to an understanding
  of the current operators at higher orders in $\kappa$.
\item {\bf Continuum models.} The algebraic construction of Section
  \ref{sec:algebraic} is worked out only for lattice models. It is
  known that certain continuum models can be obtained through special
  scaling limits of the spin chains. Such models include QFT's or the
  non-relativistic Lieb-Liniger model. It is natural to expect that
  the construction of current operators on the lattice is  relevant
  also in the scaling limit. However, this has not yet been
  investigated.
\item {\bf $T \bar T$-deformations.} It was shown in the parallel
  works \cite{sajat-ttbar,sfondrini-ttbar}  that the current operators
  also appear in the so-called bi-local deformations of the spin
  chains. These deformations can be considered as analogs of the
  famous $T\bar T$-deformation in QFT. It is an open question whether
  the actual  $T\bar T$-deformation can be established on the lattice,
  and what role the algebraic construction of \ref{sec:algebraic} can
  play in this.
\item {\bf Correlation functions in nested systems.} The current
  operators are very special: their mean values take a simple form
  even in nested spin chains, as discussed in
  \cite{sajat-longcurrents}. Local correlation functions in nested
  models have been an object of interest lately
  \cite{kluemper-su3,boos-artur-qkz-su3}. It is now generally believed
  that factorization of correlation functions can not hold generally
  in models with higher rank symmetries \cite{Smirnov-Martin-difficulties}.
  It is an interesting open question whether factorized formulas can
  be obtained at least for a subset of local operators; if the answer
  is yes, then the current operators (and their algebraic
  construction) constitute a promising starting point for future studies. 
\end{itemize}

We believe these questions deserve study, and hope that some of them will be indeed answered.

\vspace{1cm}
{\bf Acknowledgments} 

We are thankful to
Bruno Bertini, Benjamin Doyon, Frank G\"ohmann, Yunfeng Jiang,  M\'arton Kormos, Lorenzo Piroli, Herbert Spohn, G\'abor
Tak\'acs  and Eric Vernier  
for useful discussions on various questions related to the material presented here.

\appendix

\section{Current mean values using strings -- Finite volume}

\label{sec:reduced-gaudin}

Here we present the complete proof for the current
  formula (\ref{eq:xxx_finite_current_formula}) with the string solutions. As mentioned in the
  main text the key idea is to consider the divergent expressions \eqref{eq:divergences} in
  the Gaudin matrix \eqref{eq:gaudin_strings}. We introduce the transformations that reduce the
  number of these elements without altering the determinant and the
  co-factors. Then we factor out the divergences and treat the
  remaining expression.

Let us introduce the transformations $R_r$ and $L_r$ adding the $r$-th column in a given block to the $(r+1)$-th (right) and $(r-1)$-th (left) respectively. Similarly $U_r$ and $D_r$ does the same with rows (up and down). They will be used to describe the elimination of divergent expressions.

In the case of the Gaudin determinant we use the following transformation in every ($j,\varrho$) block:
\begin{equation}
    D_1 D_2 ... D_{j-1} R_1 R_2 ... R_{j-1}.
\end{equation}
After this the divergent expressions are only present in the diagonal, one in each position except the bottom-right of the blocks. These can be factored out in the leading order term (other terms do not contribute in the thermodynamic limit). What is left is the sub-determinant of bottom-right elements from all blocks. These elements were embedded in the whole matrix in such a way that no additional signs appear. Due to our transformations they contain the sum of their block meaning that we obtained exactly the reduced Gaudin determinant:
\begin{equation}
    \det G = \Bigg( \prod_{(j,\varrho} \prod_{r=1}^{j-1} K_{j,\varrho}^{r,r+1} \Bigg) \cdot \det G^{(\text{r})} \cdot \Big( 1 + \mathcal{O}(1/K) \Big).
\end{equation}

Moving on to the co-factors we have one row and column missing from the matrix. We consider three separate cases: diagonal co-factors, off-diagonal co-factors of diagonal blocks and co-factors of off-diagonal blocks. The indices for the transformations are given with respect to the situation before deleting the elements.

In the case of diagonal co-factors the corresponding diagonal block $(j,\varrho)$ is, in general, divided into four regions from which the upper-left and the bottom-right contains divergent elements. Respectively we use the transformations:
\begin{gather}
    D_1 D_2 \cdot ...\cdot D_{r-2} \cdot R_1 R_2 \cdot... \cdot R_{r-2} \\
    U_{j} U_{j-1} \cdot .. \cdot U_{r+2} \cdot L_j L_{j-1} \cdot ... \cdot L_{r+2}.
\end{gather}
where $r$ is the index of missing row and column inside the block. As a result divergent expressions are now present only in the diagonal of the block, one in each position (no exception now because one row and column is missing). In the other diagonal blocks we use the same transformation as for the determinant so that we can factor out the divergences again. The remaining factor is exactly the determinant of the reduced matrix missing one column and row, that is, the corresponding minor of the reduced matrix. It is embedded in the whole matrix properly again, implying no extra sign. The co-factor sign in the ordinary and reduced matrix match as well, since the block and the element were both diagonal. This means that for any diagonal co-factor we have
\begin{equation}
    C_{(j,\varrho, r),(j,\varrho,r)} = \Bigg( \prod_{(j,\varrho)} \prod_{r=1}^{j-1} K_{j\varrho}^{r,r+1} \Bigg) \cdot C_{(j,\varrho),(j,\varrho)}^{(\text{r})} \cdot \Big( 1 + \mathcal{O}(1/K) \Big) \hspace{1 cm} \forall r.
\end{equation}
Next we treat the off-diagonal co-factors of diagonal blocks. Let the
index (inside the block) be $r$ for the missing row and $s$ for the
column. Due to symmetry it is enough to consider $s>r$. Three of the
four regions contain divergences: the upper-left, the bottom-left and
the bottom-right. The transformations respectively are the following
(the order matters in the sense that the third must come last): 
\begin{gather}
    D_1 D_2 \cdot ... \cdot D_{r-2} \cdot R_1 R_2 \cdot ... \cdot R_{r-1} \\
    U_{j} U_{j-1} \cdot ... \cdot U_{s+1} \cdot L_j L_{j-1} \cdot ... \cdot L_{s+2} \\
    U_s U_{s-1} \cdot ... \cdot U_{r+2} \cdot R_r R_{r+1} \cdot ... \cdot R_{s-2}.
\end{gather}
The result is the same as before except that $(s-r)$ elements in the bottom-right region are present with negative sign. Nevertheless we treat the other diagonal blocks as usual and factor out again to obtain the corresponding minor of the reduced matrix. Regarding the co-factor signs the block is positive being diagonal while the element acquires the factor $(-1)^{s+r}$ which compensates for the bottom-right region. This implies that all co-factors in a diagonal block are identical:
\begin{equation}
    C_{(j,\varrho, r),(j,\varrho,s)} = \Bigg( \prod_{(j,\varrho)} \prod_{r=1}^{j-1} K_{j\varrho}^{r,r+1} \Bigg) \cdot C_{(j,\varrho),(j,\varrho)}^{(\text{r})} \cdot \Big( 1 + \mathcal{O}(1/K) \Big) \hspace{1 cm} \forall r,s.
\end{equation}
Finally let us turn our attention to the off-diagonal blocks. Let us denote the block coordinates with $(\varrho,\sigma)$. The missing row ($r$-th in block, $a$-th is matrix) and column ($s$-th in block, $b$-th is matrix) now affects two different diagonal blocks. Due to symmetry we only consider $a<b$ and $\varrho<\sigma$. The transformation for the diagonal block lacking a row is
\begin{gather}
    D_1 D_2 \cdot ... \cdot D_{r-2} \cdot R_1 R_2 \cdot ... \cdot R_{r-1} \\
    U_j U_{j-1} \cdot ... \cdot U_{r+2} \cdot L_j L_{j-1} \cdot ... \cdot L_{r+1}.
\end{gather}
For the other block simply switch $D\leftrightarrow R$ and $U\leftrightarrow L$. The rest we treat by the usual method but now serious problems arise.

Between the two deleted diagonal elements (bottom-left region of the matrix) the new diagonal is now directly below the former one. The same is true for the blocks: the ones so far below the diagonal become the new diagonal ones. As a result the divergent elements and block sums are embedded in the matrix a nontrivial way. We can still factor out the divergences to obtain the reduced Gaudin determinant but the sign of the result is unclear. We also have the co-factor signs which are $(-1)^{p+q}$ for the ordinary matrix and $(-1)^{\pi+\rho}$ for the reduced one. We show that all these signs cancel each other.

As a first step we argue that all co-factors in a given block are identical. This is clear for the absolute values so we investigate the signs. Assume we know the sign for some element $(p,q)$. Now considering one of its neighbors from the same block essentially shifts the deleted row or column. But that only has the effect of also shifting the corresponding column or row containing blocks sums. This can be reversed by a column or row interchange in the determinant implying a minus sign which exactly cancels the one appearing in the co-factor due to considering a neighboring element. This verifies our statement.

Now we look for the common sign of the block, meaning that considering a well-chosen element is enough. This should be the bottom-right one which is favorable since it causes all block sums to accumulate also in the bottom-right element (their position in the $\varrho$-th block column and $\sigma$-th block row depends on the concrete element).

To achieve a proper embedding in the bottom-left region of the matrix we move the rows containing the block sums from the bottom to the top of the given block. Applying this to the block rows $\varrho+1,\ \varrho+2,\ ...\ \sigma$ places the divergent elements and the diagonal block sums to the diagonal thus produces a correct embedding. Relocating the rows is achieved by the interchange of neighboring ones inducing minus signs. For the $\tau$-th block row with $n_\tau$ rows this produces $(-1)^{n_\tau-1}$. For the total number of row exchanges we have
\begin{equation}
    \sum_{\tau=\varrho + 1}^\sigma (n_\tau - 1) = \sum_{\tau=\varrho + 1}^\sigma n_\tau - ( \sigma - \varrho ) = (q-p) - (\sigma-\varrho).
\end{equation}
This implies the final sign
\begin{equation}
    (-1)^{q+p} \cdot (-1)^{\sigma + \varrho}
\end{equation}
compensating for the co-factor sign of the ordinary and the reduced matrix respectively. Having all sign problems eliminated we have the general identity
\begin{equation}
    C_{(j,\varrho,r),(k,\sigma,s)} = \Bigg( \prod_{(j,\varrho)} \prod_{r=1}^{j-1} K_{j,\varrho}^{r,r+1} \Bigg) \cdot C_{(j,\varrho),(k,\sigma)}^{(\text{r})} \cdot \Big( 1 + \mathcal{O}(1/K) \Big) \hspace{1 cm} \forall r,s
\end{equation}
And so the proof is complete.

\section{Decoupling of the strings}

\label{sec:decoupling}

Here we present the calculations that disentangle the string types in
the various integral equations discussed in the main text, and derive
the generating functions for the charge and current mean values.

First we complete the $a_j$ family (\ref{eq:a_functions}) defining
\begin{equation}
    a_0 (\lambda) = 2\pi \delta(\lambda),
  \end{equation}
  where now $\delta(\lambda)$ is the Dirac-delta.
  
Then we have the general convolution rule (now for $j,k\in \mathbb{N}$)
\begin{equation}
    (a_j * a_k) (\lambda) = \int_{-\infty}^\infty \frac{d\omega}{2\pi}\ a_j(\lambda-\omega) a_k(\omega) = a_{j+k} (\lambda),
\end{equation}
which can be used to prove the identities ($j>1, k\geq 1$)
\begin{align}
    a_1 * (\varphi_{j-1,k} + \varphi_{j+1,k}) - (a_0+a_2)*\varphi_{j,k} &= a_1 ( \delta_{j-1,k} + \delta_{j+1,k} ) \label{eq:convolution_identity1} \\
    a_1 * \varphi_{2,k} - (a_0+a_2)*\varphi_{1,k} &= a_1 \delta_{2,k} \label{eq:convolution_identity2}
\end{align}
that give our main tools for canceling infinite sums.
Regarding the Fourier transforms we have 
\begin{equation}
    \tilde{a}_j (p) = \int_{-\infty}^\infty \frac{d\lambda}{2\pi}\ a_j(\lambda) e^{i\lambda p} = e^{-\frac{j\abs{p}}{2}}.
\end{equation}

In the thermodynamic limit the Bethe equations become \eqref{eq:xxx_tdl_bethe_eq}
\begin{equation}
    \rho_j^t (\lambda) = \frac{1}{2\pi} a_j(\lambda) + \sum_{k=1}^\infty \int_{-\infty}^\infty \frac{d\omega}{2\pi} \varphi_{j,k}(\lambda-\omega) \rho_k(\omega).
\end{equation}
Denoting the general equation by $[j]$ we take the following combinations:
\begin{equation}
    \begin{cases}
        (a_0 + a_2)* [j] - a_1* \Big( [j-1] + [j+1] \Big) & j>1 \\
        (2a_0+a_2) * [1] - a_1 * [2] & j=1
    \end{cases}
    \label{eq:decoupling_trick}
\end{equation}
so that our convolution identities (\ref{eq:convolution_identity1}) and (\ref{eq:convolution_identity2}) can be utilized. We obtain the equations
\begin{equation}
    \begin{cases}
        (a_0 + a_2) * \rho_j^t = a_1 * (\rho_{j-1}^h + \rho_{j+1}^h) & j>1 \\
        (a_0 + a_2) * \rho_1^t = a_1 + a_1*\rho_2^h & j=1,
    \end{cases}
\end{equation}
which can be simplified most conveniently in Fourier space. The inverse transformation can be performed using the identity
\begin{equation}
    \int_{-\infty}^\infty dp\ \frac{1}{2\cosh{\frac{p}{2}}} e^{-i\lambda p} = \frac{\pi}{\cosh{\pi\lambda}} \eqqcolon s(\lambda)
\end{equation}
leading to the final general result given by eq. \eqref{eq:decoupled_bethe_eqs} in the main text.

Applying these decoupled equations one can simplify both the charge and the current mean values as it is shown below. During our derivation we will use the identity 
\begin{equation}
    s* (a_{j-1} + a_{j+1}) = a_j \hspace{0.7 cm} j \geq 1
    \label{eq:s_convolution_identity}
\end{equation}
which is easily checked in Fourier space.

For the charge mean value
\begin{equation}
    \frac{ \expval{Q(\nu)}}{L}
    = \sum_{j=1}^\infty \int_{-\infty}^\infty d\lambda\ \rho_j(\lambda) h_{j}(\lambda|\nu)
\end{equation}
we substitute $h_{j}$ from (\ref{eq:string_charges}) and
express the particle density using the decoupled equations
(\ref{eq:decoupled_bethe_eqs}):
\begin{equation}
    \rho_j = \rho_j^t-\rho_j^h = \frac{1}{2\pi} \delta_{j,1}s + s * (\rho_{j-1}^h+\rho_{j+1}^h) - \rho_j^h
\end{equation}
to acquire the form
\begin{align}
    \frac{\expval{Q(\nu)}}{L} = - &\Bigg[ s*a_1 + 2\pi \rho_1^h * ( s * a_2 - a_1 ) + \nonumber \\
    &+ 2\pi \sum_{j=2}^\infty \rho_j^h * \Big( s* (a_{j-1} + a_{j+1}) - a_j) \Big) \Bigg] (\nu) 
\end{align}
Applying (\ref{eq:s_convolution_identity}) we see that the infinite sum cancels term-by-term while the other part gives the result \eqref{eq:xxx_tdl_charge_meanvalue}.

For the currents we start from the dressing equations
\begin{equation}
    x_j = x_j^\text{dr} - \sum_{k=1}^\infty \varphi_{j,k} * (f_k x_k^\text{dr})
\end{equation}
and take their combination according to
(\ref{eq:decoupling_trick}). These can be simplified again by using
the convolution identities (\ref{eq:convolution_identity1}) and
(\ref{eq:convolution_identity2}) first then going to Fourier space. After
the inverse transformation we have 
\begin{equation}
    \begin{cases}
        x_j - s*(x_{j-1} + x_{j+1}) = x_j^\text{dr} - s* \Big( \eta_{j-1} x_{j-1}^\text{dr} + \eta_{j+1} x_{j+1}^\text{dr} \Big) & j>1 \\
        x_1 - s*x_2 = x_1^\text{dr} - s*(\eta_2 x_2^\text{dr}) & j=1,
    \end{cases}
\end{equation}
where we defined $\eta_j = \rho_j^h/\rho_j^t$. So far we worked with
the general functions $x_j$ but since the current formula contains the
charge eigenvalues let us now set
\begin{equation}
    x_j (\lambda) \coloneqq h_{j}(\lambda|\nu) = - a_{j}(\lambda-\nu).
\end{equation}
This makes the left hand side vanish for $j>1$ and creates the source term for $j=1$, according to (\ref{eq:s_convolution_identity}). Thus we obtain \eqref{eq:decoupled_dresing_equations-altalanos}.

Now we use this to reformulate the current formula. Substituting the string charges (\ref{eq:string_charges}) we have
\begin{gather}
    \expval{J(\mu,\nu)} = \sum_{j=1}^\infty \int_{-\infty}^\infty \frac{d\lambda}{2\pi}\ f_j(\lambda) h_{j}'(\lambda|\nu) h_{j}^\text{dr}(\lambda|\mu) = \\
    = \partial_\nu \Bigg[ \sum_{j=1}^\infty a_j * \Big( h_{j}^\text{dr}(\cdot|\mu) - \eta_j h_{j}^\text{dr}(\cdot|\mu) \Big) \Bigg] (\nu).
  \end{gather}
Here it is understood that the convolution acts on the first variable
of the functions $h_{j}^\text{dr}$.

We substitute $h_j^\text{dr}$ from
\eqref{eq:decoupled_dresing_equations-altalanos}, but only for the
first term, and then rearrange the sum to obtain
\begin{align}
    \expval{J(\mu,\nu)} = \partial_\nu \Bigg[ - \frac{1}{2\pi} &\Big( a_1 * s \Big)(\nu-\mu) + \bigg( a_2 * s * \big( \eta_1 h_{1}^\text{dr}(\cdot|\mu) \big) - a_1 * \big( \eta_1 h_{1}^\text{dr}(\cdot|\mu) \big) + \nonumber \\
    &+ \sum_{j=2}^\infty \Big[ s * ( a_{j-1} + a_{j+1} ) - a_j \Big] * \big( \eta_j h_{j}^\text{dr}(\cdot|\mu) \big) \bigg)(\nu) \Bigg].
\end{align}
According to (\ref{eq:s_convolution_identity}) the sum cancels term-by-term again while the rest gives \eqref{eq:xxx_tdl_current_formula}.

\section{Hydrodynamic projections and asymptotic correlation functions}

\label{sec:proj}

Here we give some details about the computation of the amplitudes describing the asymptotic behaviour of dynamical correlation
functions. We do not discuss the theory of hydrodynamic projections, instead we refer the reader to
\cite{Doyon-GHD-LM,doyon-hydro-projection}. Below we use the formalism of \cite{Doyon-GHD-LM}.

We start with the average of local operators in a GGE, described by some Lagrange multipliers $\beta_\alpha$. 
The hydrodynamic projection to the $\alpha$-th charge of an observable $\mathcal{O}$ is given by the derivative of the
mean values with respect to the corresponding Lagrange multiplier. 
One can always define suitable spectral functions $V_j^\mathcal{O}(\lambda)$ such that the projection is of the form
\begin{equation}
 \frac{\partial}{\partial \beta_\alpha} \expval{\mathcal{O}} = \sum_{j=1}^\infty \int_{-\infty}^\infty d\lambda\ \rho_j^t(\lambda) \Big( \partial_{\beta_\alpha} f_j(\lambda) \Big) V_j^{\mathcal{O}}(\lambda).
    \label{eq:xxx_Vfactor_def}
\end{equation}

As mentioned in the main text the building blocks have a simple relation to the generalized current mean values:
\begin{equation}
    \expval{J_{\alpha,\gamma}} = \frac{1}{2} \Psi_{\gamma-1,\alpha-2}.
\end{equation}
Therefore we only need the derivatives of the currents according to (\ref{eq:xxx_Vfactor_def}) and the desired amplitudes $V_j^{\Psi_{\alpha,\delta}}$ can be simply read off. In our derivation we use the differentiated dressing equation
\begin{equation}
    \partial_{\beta_\delta} x_j^\text{dr} = \sum_{k=1}^\infty \varphi_{j,k}* \bigg( (\partial_{\beta_\delta} f_k) x_k^\text{dr} + f_k (\partial_{\beta_\delta} x_k^\text{dr}) \bigg).
    \label{eq:differentiated_dressing_eq}
\end{equation}
We analyze a general expression with any spectral functions $\{x_j\}_{j=0}^\infty$ and $\{y_j\}_{j=0}^\infty$. For simplicity we hide the $\lambda$ dependence of all functions.
\begin{equation}
    \partial_{\beta_\delta} \int_{-\infty}^\infty \frac{d\lambda}{2\pi}\ x_j f_j y_j^\text{dr} = \int_{-\infty}^\infty \frac{d\lambda}{2\pi}\ x_j \bigg( ( \partial_{\beta_\delta} f_j ) y_j^\text{dr} + f_j ( \partial_{\beta_\delta} y_j^\text{dr} ) \bigg).
\end{equation}
According to (\ref{eq:differentiated_dressing_eq}) we successively substitute $\partial_{\beta_\delta} y_j^\text{dr}$ obtaining an infinite series
\begin{equation}
   \int_{-\infty}^\infty \frac{d\lambda}{2\pi} \bigg( x_j ( \partial_{\beta_\delta} f_j ) y_j^\text{dr} + x_j f_j \sum_{k=1}^\infty \varphi_{j,k} * \Big( ( \partial_{\beta_\delta} f_k ) y_k^\text{dr} \Big) +\ ... \ \bigg).
\end{equation}
Using the operator $\underline{\underline{\hat{T}}}$ and the compact formalism defined in (\ref{eq:xxx_dressing_equation_compact}) (now omitting the matrix and vector notations) we have
\begin{gather}
   \int_{-\infty}^\infty \frac{d\lambda}{2\pi}\ x_j f_j \Bigg[ \Bigg( \sum_{k=0}^\infty \hat{T}^k \Bigg) \bigg( \frac{\partial_{\beta_\delta} f}{f} y^\text{dr} \bigg) \Bigg]_j = \\
    = \int_{-\infty}^\infty \frac{d\lambda}{2\pi}\ x_j f_j \Bigg[ \Big(1 - \hat{T} \Big)^{-1} \bigg( \frac{\partial_{\beta_\delta} f}{f} y^\text{dr} \bigg) \Bigg]_j,
\end{gather}
where the dressing operator can be recognized. We use its symmetry to obtain our final result
\begin{equation}
    \int_{-\infty}^\infty \frac{d\lambda}{2\pi}\ (\partial_{\beta_\delta} f_j) x_j^\text{dr} y_j^\text{dr}.
\end{equation}
Simple substitutions lead to the derivative of the current mean value:
\begin{equation}
    \partial_{\beta_\delta} \expval{J_{\alpha,\gamma}} = \sum_{j=1}^\infty \int_{-\infty}^\infty \frac{d\lambda}{2\pi}\ \Big( \partial_{\beta_\delta} f_j(\lambda) \Big) \big( h_{\gamma,j}' \big)^\text{dr}(\lambda) h_{\alpha,j}^\text{dr}(\lambda)
\end{equation}
from which the desired functions for the building blocks can be read off according to \eqref{eq:xxx_Vfactor_def}. Thus we obtain \eqref{eq:building_block_Vfactor}.


\providecommand{\href}[2]{#2}\begingroup\raggedright\endgroup

\end{document}